\documentclass[%
 reprint,
 amsmath,amssymb,
 aps,
]{revtex4-2}

\usepackage{booktabs}
\usepackage{graphicx}
\usepackage{dcolumn}
\usepackage{bm}
\usepackage[version=3]{mhchem}
\usepackage{pgfplots}
\usepackage{tikz}
\usepgfplotslibrary{groupplots}
\pgfplotsset{compat=1.18}
\usepackage{tabularx} 
\usepackage{booktabs}
\usepackage{amsmath}
\usepackage{subcaption}
\usepackage{stackengine}
\usepackage{dblfloatfix}
\usepackage{placeins}
\usepackage{float}
\usepackage{soul}
\usepackage{xcolor}
\usepackage{soulutf8}
\usepackage{url}

\sethlcolor{cyan}

\begin{document}

\preprint{}

\title{Partially reactive force field for the UiO-66 metal-organic framework}

\author{Akanksha Nawani}
\author{Rocio Semino}%
 \email{rocio.semino@sorbonne-universite.fr}
\affiliation{Sorbonne Université, CNRS, Physicochimie des Electrolytes et Nanosystèmes Interfaciaux, PHENIX, F-75005 Paris, France.}%

\date{\today}

\begin{abstract}
UiO-66 is the most widely studied metal-organic framework, on account of its structural tunability given by its capacity of sustaining high amounts of point defects in its structure. Its synthesis mechanism is largely unknown, with only a few works mostly focused on the formation of the Zr-oxide cluster. In this work, a partially reactive force field to model UiO-66, nb-UiO-FF, is introduced. This force field incorporates node--ligand reactivity via a Morse potential and the introduction of dummy atoms to reproduce the anisotropic charge distribution of the Zr atoms in the node. nb-UiO-FF reproduces structural features of both UiO-66 and its isoreticular analog UiO-67 as well as framework stability with or without defects, activated or filled with N,N-dimethylformamide or ethanol. Defect generation and node--ligand binding energies are comparable with density functional theory values. The force field is further employed within a molecular dynamics scheme to study the early stages of solvothermal node--ligand binding. Species with structures corresponding to motifs present in UiO-66 are observed, along with other metastable species. This force field  enables studying the self-assembly of UiO-66, as well as the formation of its point defects.  
\end{abstract}


\maketitle


\section{Introduction}

Metal-organic frameworks (MOFs) are a popular class of porous materials formed by the coordination of metal ions or oxoclusters with organic ligands.\cite{Hoskins1989,Kondo1997,Yaghi1995,Moghadam2017} Their chemical and physical properties can be easily modified by changing their constituent building blocks, which can include a wide variety of ligands and metals across the periodic table, leading to different framework topologies. Their tunable functionality, high surface areas and customizable pore architecture make them promising candidates for diverse applications including gas storage, drug delivery and catalysis, among others.\cite{Sahayaraj2023,Andrade2023}

\begin{figure}[htbp]
    \centering
    \noindent
    \begin{minipage}[b]{0.38\columnwidth} 
        \centering
        \begin{subfigure}{\linewidth}
            \raggedright (a) \\ 
            \centering
            \includegraphics[trim=15px 10px 20px 10px, clip, width=0.7\linewidth]{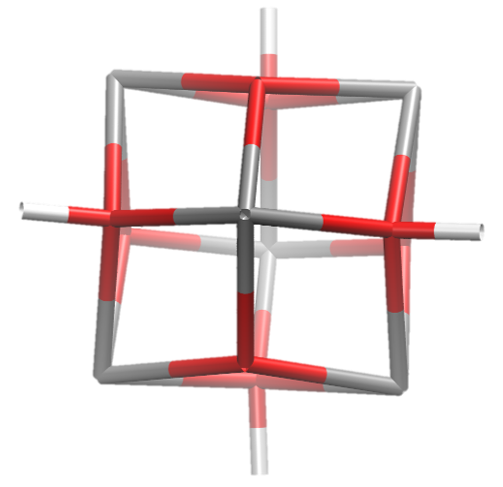}
        \end{subfigure}\\[1ex]
        \begin{subfigure}{\linewidth}
            \raggedright (b) \\
            \centering
            \includegraphics[trim=150px 100px 150px 100px, clip, width=0.7\linewidth]{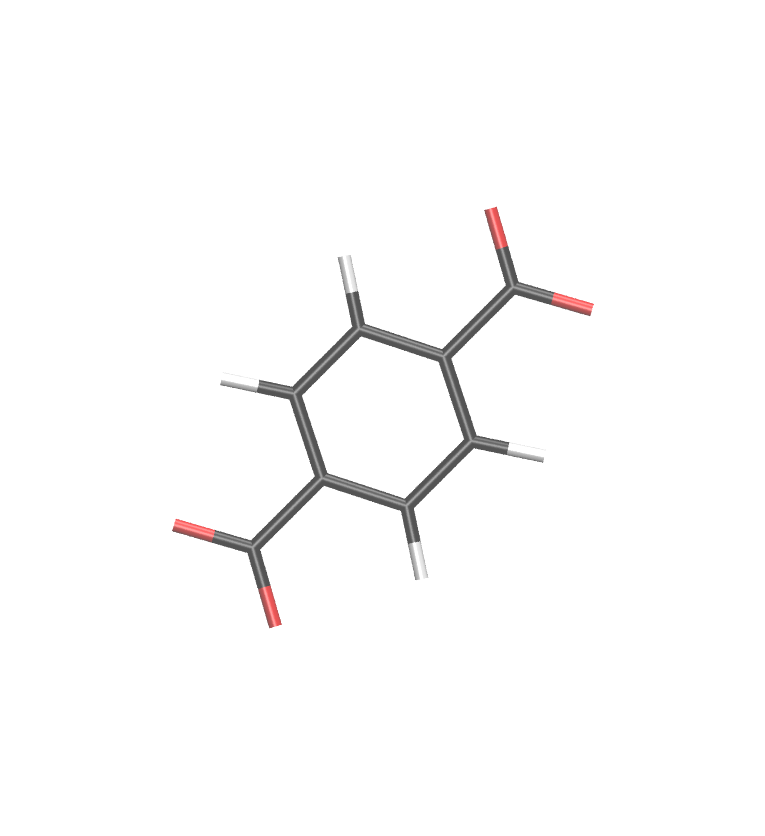}
        \end{subfigure}
    \end{minipage}%
    \hfill
    \begin{minipage}[b]{0.62\columnwidth} 
        \centering
        \begin{subfigure}{\linewidth}
            \raggedright (c) \\
            \centering
            \includegraphics[trim=40px 10px 40px 10px, clip, width=\linewidth, height=6cm, keepaspectratio]{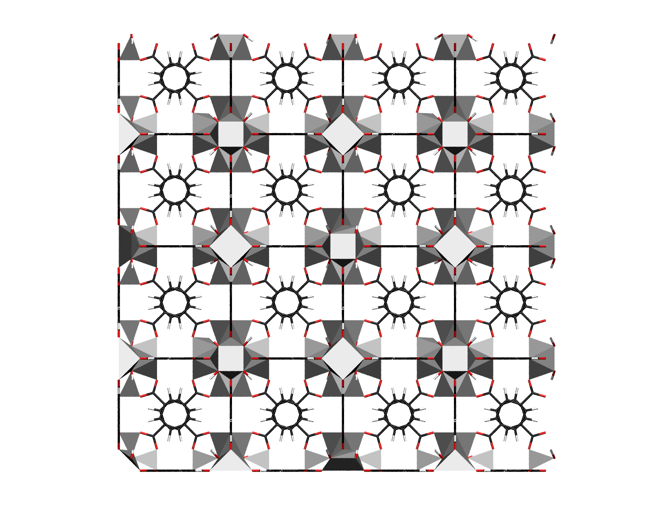}
        \end{subfigure}
    \end{minipage}
    \caption{Structural representation of UiO-66: (a) $Zr_6(\mu_3-O)_4(\mu_3-OH)_4$ metal cluster, (b) Benzene-1,4-dicarboxylate (BDC) linker, and (c) a 2$\times$2$\times$2 supercell of  UiO-66 with the metal cluster represented as polyhedra. Color code: O (red), H (white), Zr (silver), C (gray)}
    \label{fig:uio}
\end{figure}

UiO-66\cite{Cavka2008} is one of the most studied MOFs, owing to its high stability and widespread applications.\cite{Dhakshinamoorthy2018, vermoortele2013, woellner2018} It is composed of Zr-oxide metal nodes and benzene dicarboxylic acid (BDC) ligands bound together following the \texttt{fcu} topology, as shown in Fig.~\ref{fig:uio}. It possesses high thermal,\cite{Athar2021} chemical\cite{Cavka2008, Ahmadijokani2020} and mechanical stability,\cite{Wu2013,Yot2015,Cavka2008} which arise from the strong coordination bonds between the \ce{Zr} and \ce{O} atoms of the BDC ligand carboxylate group.\cite{Cavka2008} Additionally, the high coordination number of \ce{Zr} provides the ability to withstand multiple point defects, including missing ligand and missing node, offering a great arena for defect engineering,\cite{DeStefano2017,Feng2019,Cox2023} with an impact on adsorption properties.\cite{Liang2016} UiO-66 exists in two forms: hydroxylated and dehydroxylated, which can interchange through a reversible process that involves the loss of two water molecules per metal cluster.\cite{Valenzano2011}

With the emergence of computational techniques to study materials, many computational works have been dedicated to understand the properties of MOFs and facilitate their discovery for target applications.\cite{Demir2023,Moghadam2024,Temmerman2024} Among the techniques used, force field-based simulations have been instrumental in the field, as they enable to cover systems with sizes and timescales compatible with guest diffusion and guest-induced phase transitions.\cite{Chen2022} These simulations require having a force field to properly model the physicochemical behaviour of the MOF and its interactions with guest molecules. Generalized force fields like UFF4MOF\cite{Coupry2016} and specific MOF force field development methods, including BTW-FF,\cite{Bristow2014} MOF-FF \cite{Bureekaew2013} and Quick-FF\cite{Vanduyfhuys2015} were devised.\cite{Heinen2018} However, most force fields struggle to model the hydroxylated version of UiO-66, due to the complex interplay between intra-cluster and cluster-ligand interactions.\cite{boyd2017} Most importantly, these force fields are unable to capture reactive processes, since they operate over a pre-defined connectivity. This renders them unable to model MOF and defect formation.

In order to model MOF reactivity, modern machine learning force fields have been developed.\cite{Dobbelaere2025} However, developing accurate machine learning force fields for MOFs remains challenging,\cite{Tayfuroglu2026} and modelling both defect formation and self-assembly as a function of synthesis conditions requires being able to easily expand the force field to include other species, such as solvents, counterions and modulators.  As an alternative, the Reax-FF\cite{Senftle2016} force field may be used, but its validity for accurately modelling MOF properties is still under discussion.\cite{Zhuo2026,Castel2022,Castel2023} Biswal and Kusalik introduced the use of cationic dummy atom (CDA) models to capture metal--ligand interactions in solution,\cite{Biswal2016} which was later on complemented by modeling dispersion interactions via a Morse term by Balestra and Semino for ZIFs.\cite{Balestra2022} These force fields were successfully used to study the self-assembly of MOF-2\cite{Biswal2016,Biswal2017}  and ZIF-8,\cite{AndarziGargari2026,AndarziGargari2025,Balestra2022} as well as for modelling phase transformations and metal--ligand binding events for the ZIF-4 MOF.\cite{Mendez2024,Mendez2024_2,Mendez2025} CDA force fields were also developed for MOFs that contain metal oxoclusters as nodes as well.\cite{Jawahery2019,Su2021,Golo2025} In particular, Su and Ahlquist developed a CDA model for UiO-66,\cite{Su2021} and have shown that it reproduces non-defective framework stability in a series of solvents. In their model, each \ce{Zr} atom contains eight dummy particles, which implies a total of 48 dummy atoms per node, along with the possibility of modelling node decomposition. 

In this work, we develop a CDA force field for UiO-66 (nb-UiO-FF, where \texttt{nb} stands for non-bonded) that accurately models its structural, dynamic and reactive properties. In addition of being transferable to its isoreticular variant UiO-67, our force field allows to model UiO-66 frameworks containing point defects, including missing ligand and missing node, while maintaining framework integrity. Our CDA model contains four dummy atoms per \ce{Zr} atom, increasing model efficiency with respect to that from Su and Ahlquist at the expense of losing intranode reactivity. Crucially, nb-UiO-FF can be applied to studying its self-assembly, defect formation and stability, as it incorporates node--ligand reactivity.

This article is organized as follows. The concept and methodology followed for developing the nb-UiO-FF force field are presented in Section II. Section III covers the validation of the force field and results for modelling early self-assembly stages. Section IV summarizes our conclusions. 

\section{Methods} 

\subsection{Force field development}

This study builds on the methodology previously developed within our group to model ZIF-8.\cite{Balestra2022} Within this approach, the hard, bonded potentials between metal node and ligand are replaced with a non-bonded Morse potential and the incorporation of CDA. This framework allows for the simulation of dynamic bond breaking and formation while ensuring that the correct bond topology is obtained, which is essential for investigating reactive processes.

All the simulations were carried out using the LAMMPS simulation package.\cite{lammps} The initial coordinates for UiO-66 were taken from the study of Boyd \textit{et al.}\cite{boyd2017} Our parameterization takes as starting point the non-reactive force field developed by Yang and coworkers for studying gas diffusivity in UiO-66.\cite{yang2011} However, this force field was designed for the dehydroxylated structure, characterized by a \( \mathrm{Zr_6O_6} \) metal node, whereas our focus is on the more common hydroxylated form (\ce{Zr6O4(OH)4}), in which the composition of the metal node differs. To account for this, we adopted the bonded force constants (bonds and angles) and non-bonded parameters for the metal cluster from the reported model, while the corresponding equilibrium bond lengths and angles were taken from Density Functional Theory (DFT) results reported by Valenzano and coworkers.\cite{Valenzano2011} For the ligand, both bonded (bond, angle, dihedral, and improper) and non-bonded parameters, including their equilibrium values, were adopted entirely from the same model.  Subsequently, all bonded interactions (bonds, angles and dihedrals) between the metal cluster and ligands were removed, to be replaced by a single Morse potential term. 

Bond stretching and angle bending interactions were described using harmonic potentials and dihedral and improper torsions were modelled by periodic cosine potential functions. Electrostatic interactions were treated using fixed partial charges, adopted from the force field by Yang and coworkers,\cite{yang2011} with the exception of those corresponding to the metal node oxygen atoms, which were modified to reflect the hydroxylated structure. The total charge of the six oxygen atoms in the dehydroxylated model was redistributed among oxygen-containing species in the hydroxylated node, \(\mu_3\)-O and \(\mu_3\)-OH (including the hydrogen), while preserving all the other charges from the original partitioning scheme. Lennard-Jones (LJ) interactions were used to describe short-range van der Waals interactions within a cutoff of 14 \AA. Long-range electrostatic interactions were computed using the Ewald summation method under periodic boundary conditions.
The functional form of the force field is given in equation \ref{eq:ff}:
\begin{equation}
\begin{aligned}
U_{\text{nb-UiO-FF}} &= \sum_{\text{bonds}} k_r (r - r_0)^2 \\
&+ \sum_{\text{angles}} k_\theta (\theta - \theta_0)^2 \\
&+ \sum_{\text{dihedrals}} k_\phi \left[1 + \cos(n\phi - \delta)\right] \\
&+ \sum_{\text{impropers}} k_\omega \left[1 + \cos(n\omega - \delta)\right] \\
&+ \sum_{i<j} \left[ \frac{q_i q_j}{4\pi\varepsilon_0 r_{ij}} + U_{\text{vdW}}(r_{ij}) \right]
\end{aligned}
\label{eq:ff}
\end{equation}

\noindent
where the van der Waals interactions are described by the Lennard--Jones potential for all atom pairs except the $Zr-O_{\mathrm{carboxyl}}$ bonds, which are described using a Morse potential instead:

\begin{equation}
\displaystyle
U_{\mathrm{vdW}}(r_{ij}) =
\begin{cases}
U_{\mathrm{LJ}}(r_{ij}), & \text{all pairs except } Zr-O_{\mathrm{carboxyl}},\\
U_{\mathrm{Morse}}(r), & Zr-O_{\mathrm{carboxyl}},
\end{cases}
\label{eq:vdW}
\end{equation}

and Lennard-Jones terms follow the equation: 

\begin{equation}
U_{\mathrm{LJ}}(r_{ij}) =
4\varepsilon_{ij}
\left[
\left(\frac{\sigma_{ij}}{r_{ij}}\right)^{12}
-
\left(\frac{\sigma_{ij}}{r_{ij}}\right)^6
\right],
\label{eq:lj}
\end{equation}

\noindent
with $r$ and $r_0$ the instantaneous and equilibrium bond lengths respectively; $\theta$ and $\theta_0$ the bond angle and equilibrium angle; $\phi$ and $\omega$ dihedral and improper torsional angles; $k_r$, $k_\theta$, $k_\phi$, and $k_\omega$ the corresponding force constants; $n$ the multiplicity, and $\delta$ the phase shift. In addition, $q_i$ and $q_j$ are the atomic partial charges, $r_{ij}$ is the interatomic distance, $\varepsilon_{ij}$ and $\sigma_{ij}$ are the LJ well depth and size parameters, respectively.

To sustain the correct topology of ligands around the metal center in the absence of bonded potentials, we adopted the CDA model. This involved introducing a dummy atom along each $Zr-O_{ligand}$ bond resulting in four dummy atoms per metal center as shown in Fig. S2. These dummy atoms (denoted as D) are connected to Zr through strong bonded potentials (bond and angle force constants of $k_{\mathrm{Zr-D}} = 800\ \text{kcal mol}^{-1}\ \text{\AA}^{-2}$ and $k_{\mathrm{D-Zr-D}} = 150\ \text{kcal mol}^{-1}\ \text{rad}^{-2}$, respectively). Electrostatic interactions between ligand and dummy atoms preserve the directionality of the metal-ligand bonds and play a key role in maintaining the correct coordination geometry.\cite{Biswal2017} Different charge distributions between the \ce{Zr} and the dummy atoms were tested; however, only configurations with very low charge on the \ce{Zr} center preserved structural stability. Hence, the charge on \ce{Zr} of +1.968 was fully distributed over its four dummy atoms, each carrying a charge of +0.492, and the \ce{Zr} atom was rendered electrically neutral.  

A Morse potential was introduced to describe the $Zr-O_{ligand}$ reactivity and is given in Equation \ref{eq:morse}: 
 \begin{equation}
  \displaystyle
    U_{\mathrm{Morse}} = D_0 \left[ e^{- 2 \alpha (r - r_0)} - 2 e^{- \alpha (r - r_0)} \right]
     \label{eq:morse}
 \end{equation}
 where {$U_{\mathrm{Morse}}$} is the potential energy. $D_0$ is the well depth, $\alpha$ controls the width of the potential well, $r$ is the instantaneous interatomic distance, and $r_0$ is the equilibrium bond distance. The parameters $D_0$ and $\alpha$ were subsequently optimized following the procedure detailed below. The equilibrium bond distance $r_0$ was taken from Valenzano \textit{et al.}\cite{Valenzano2011} to ensure consistency with the other bonded interactions. No additional dispersive interactions were included between these species. Note that the Morse depth should not be interpreted independently as the Zr--O binding energy, as the binding energy will be given by the Morse term together with electrostatic contributions from the dummy--ligand interactions.  

 \begin{figure}[htbp]
    \centering
    
    \begin{subfigure}[b]{0.48\textwidth}
        \centering
        \includegraphics[width=\textwidth]{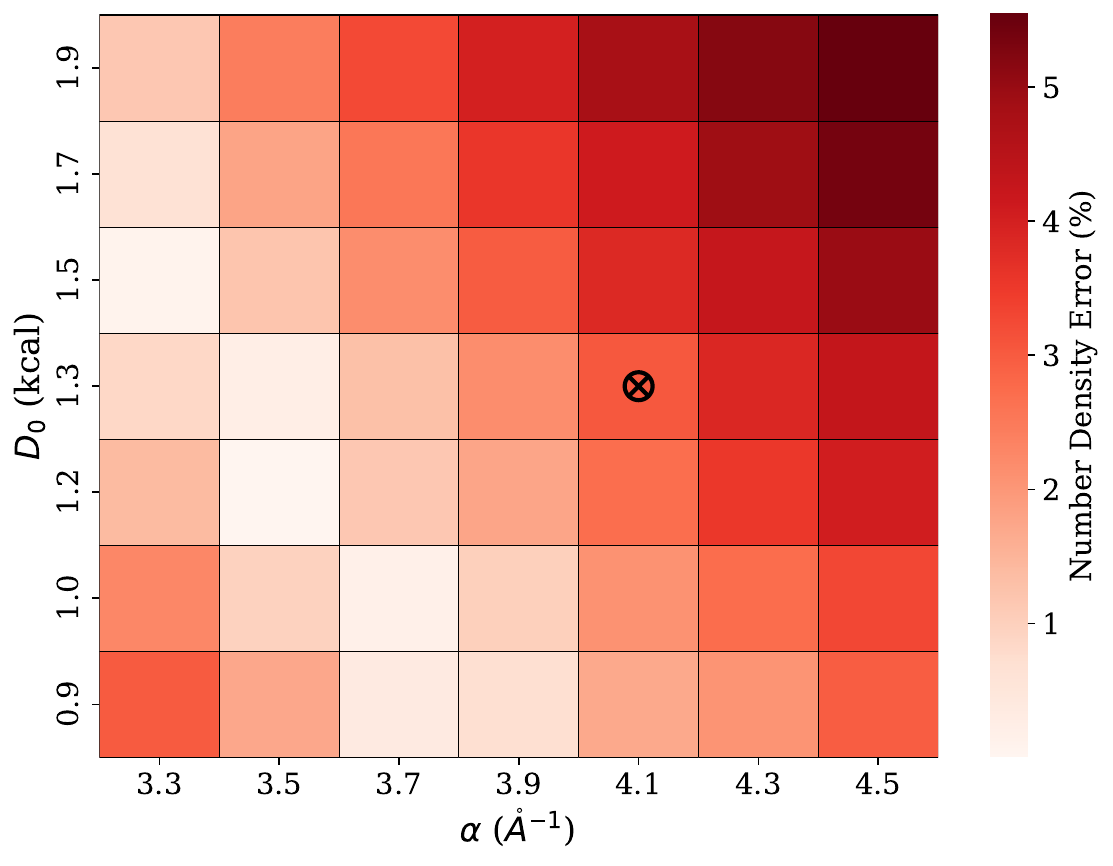}
        \subcaption{}
        \label{fig:density_error}
    \end{subfigure}
    \hfill 
    \begin{subfigure}[b]{0.48\textwidth}
        \centering
        \includegraphics[width=\textwidth]{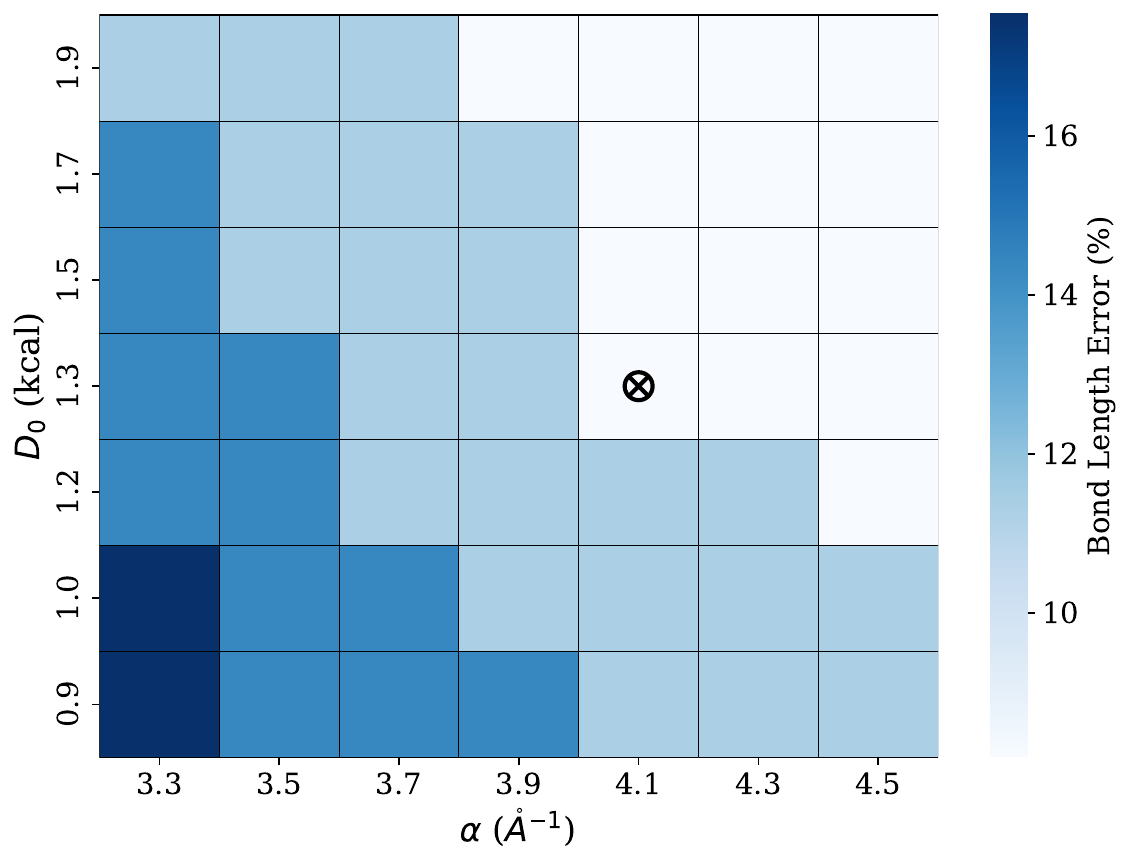}
        \subcaption{}
        \label{fig:energy}
    \end{subfigure}
    
    \caption{Parameter space exploration for the Morse potential parameters $D_0$ and $\alpha$. (a) Percentage error in the calculated number density relative to the theoretical value. (b) Percentage error in the first peak position of the $Zr-O_{\mathrm{ligand}}$ radial distribution function relative to DFT reference values from Ref.~\citenum{Valenzano2011}.}
    \label{fig:parameter_heatmaps}
\end{figure}

 Parameters optimization was carried out through a systematic exploration of $D_0$-$\alpha$ pairs. For each pair, short molecular dynamics simulations were performed for a 2$\times$2$\times$2 supercell at 298 K temperature and 1 atm pressure using an automated custom Python script. Parameters sets that resulted in structural collapse or significant distortion of the framework were discarded. For the remaining candidates, the simulated number density and $Zr-O_{ligand}$ distances were calculated and compared with experimental values, and the parameter sets yielding the lowest error were identified. A plot of the number density error ($\%$) and the percentage error in the first peak position of the $Zr-O_{ligand}$ radial distribution functions (RDF) relative to DFT values\cite{Valenzano2011} is presented in Fig.~\ref{fig:parameter_heatmaps}. While certain parameter combinations yielded marginally lower number density errors, this improvement came directly at the expense of the local atomic geometry. RDF analysis showed that as the density error was minimized, the $Zr-O_{ligand}$ bond length increasingly deviated from its actual value. The final parameters were therefore selected to provide an optimal balance, prioritizing accurate local bond lengths while keeping the number density within 5\% margin and the first RDF peak within 10\% of the reference DFT values. The selected parameter set is highlighted by a cross marker in Fig. ~\ref{fig:parameter_heatmaps} and presented in the SI, along with all other force field parameters.

\subsection{Simulation details}

The force field for dimethylformamide (DMF) was adopted from Chalaris \textit{et al,}\cite{Chalaris2000} while ethanol was modelled using the TraPPE force field\cite{Morales2013} parameters. Non-bonded cross-interactions were computed using the Lorentz--Berthelot mixing rules. In cases where the resulting LJ well depth $\epsilon$ was originally zero (e.g., for dummy atoms and certain hydrogen sites), a small finite value of $1 \times 10^{-5}$ was assigned instead to avoid numerical instabilities. For ethanol, it was observed that the hydroxyl hydrogen exhibited artificially strong interactions with the $\mu_3$-O and $\mu_3$-OH atoms of the metal cluster. To mitigate this, the Lennard-Jones $\sigma$ parameter for crossed H$_{OH_{ethanol}}$--O$_{node}$ interactions was increased by 30\%  from that computed via the Lorentz-Berthelot mixing rules (from 1.559 to 2.027~\AA), effectively introducing a short-range repulsion.

The solvents were introduced in the framework using hybrid MD and Grand Canonical Monte Carlo (GCMC) simulations \cite{frenkel2023understanding} in the $\mu VT$ ensemble. The simulations were performed at 298 K with a chemical potential corresponding to a bulk pressure of 1 atm. The MD component was carried out in the NVT ensemble using a timestep of 0.5 fs and a Nosé–Hoover\cite{Nose1984, hoover85} thermostat with a damping constant of 50 fs. GCMC moves were attempted every 500 MD steps. At each exchange step, an average of 100 insertion/deletion attempts were carried out, with equal probability assigned to insertion and deletion moves. No translational or rotational Monte Carlo moves were performed. The MD steps allowed structural relaxation and diffusion of adsorbed molecules within the porous framework between successive GCMC exchange steps. A saturation loading was reached when no additional molecules were accepted and the number of adsorbed molecules plateaued (see Fig. S6).

For the MD simulations performed to assess structural stability in the presence of solvents, the system was first subjected to energy minimization using the conjugate gradient algorithm, followed by equilibration in the NVT ensemble at 298 K for 50 ps using a timestep of 0.5 fs and a Nosé–Hoover thermostat with a  damping constant of 50 fs. The equilibrated configurations were then used as starting point for 20 ns production simulations in the NPT ensemble at 298 K and 1 atm using a Nosé–Hoover thermostat with a dampling constant of 50 fs and barostat \cite{Nose1984, hoover85} with a damping constant of 500 fs. A timestep of 0.5 fs was used. To prevent spurious center-of-mass drifts arising during the long NPT simulations, the total linear and angular momentum of the system were periodically removed every 100 timesteps.  

The same protocol and parameters described above were consistently applied to all defective systems. Simulations were initially performed without solvent, after which DMF and ethanol were introduced via GCMC simulations using identical conditions as those mentioned above. The resulting configurations were then subjected to 20 ns of molecular dynamics simulations to evaluate structural stability. The force field parameters for the capping species were adopted from UFF.\cite{UFF1992} 

Regarding the simulations performed for computing mechanical properties, the system was first equilibrated at 298 K and 1 atm in the NPT ensemble for 50 ps using Nosé–Hoover thermostat and barostat, followed by 50 ps in the NVT ensemble using a Nosé–Hoover thermostat to obtain a relaxed reference configuration. Deformations were then applied by modifying the simulation cell dimensions and remapping atomic positions accordingly. For the uniaxial deformation, the simulation cell was strained along a single direction (e.g., the $x$-axis) while keeping the other directions unchanged. Similar procedures were employed for biaxial and shear deformations, with appropriate modifications to impose the desired strain states.
Different magnitudes of strain were systematically applied to construct the energy–strain relationships (presented in Fig. S5), with multiple intermediate values sampled over symmetric ranges around zero strain. Specifically, uniaxial strains were varied between $-6\%$ and $+6\%$, biaxial strains between $-3 \%$ and $+3\%$, and shear strains between $-10\%$ and $+10\%$. Following each deformation, the system was equilibrated for 0.5 ns in the NVT ensemble at 298 K, and the potential energy and cell properties were subsequently averaged over an additional 1 ns.

Furthermore, the same simulation protocol and parameters decribed above were applied to model bulk UiO-67. Simulations were first performed in the absence of solvent, followed by GCMC-based DMF loading, after which molecular dynamics simulations were carried out for 20 ns.

For the self-assembly simulations, the systems were first energy minimized and equilibrated using a timestep of 0.5 fs. The temperature was gradually increased from 1 to 298 K in the NVT ensemble using a Nosé–Hoover thermostat with a damping constant of 50 fs. This was followed by equilibration in the NPT ensemble using Nosé–Hoover thermostat and barostat with a damping constant of 50 fs for thermostat and 500 fs for barostat at 298 K and 1 atm. Production simulations were then carried out in the NPT ensemble for a short time of 6 ns to observe the species formed in the early stages of self-assembly.

Radial Distribution Functions (RDFs) were obtained using VMD (1.9.3).\cite{HUMPHREY199633} A bin width of 0.01 \AA \ was used in all cases and periodic boundary conditions were taken into account.

\subsection{DFT calculations}

DFT calculations were performed to validate our force field with respect to energetics using the pw.x code from the open-source suite Quantum Espresso 7.3.\cite{Giannozzi2009} The PBEsol functional \cite{PBESol} was considered and dispersion corrections where obtained by applying the Grimme-D3\cite{Grimme2010} scheme. We used the following pseudopotentials from the \textsc{Quantum ESPRESSO} pseudopotential database: \texttt{Zr.pbesol-spn-kjpaw\_psl.1.0.0.UPF}, \texttt{O.pbesol-n-kjpaw\_psl.1.0.0.UPF}, \texttt{C.pbesol-n-kjpaw\_psl.1.0.0.UPF}, \texttt{H.pbesol-rrkjus\_psl.0.1.UPF} and \texttt{Cl.pbesol-n-kjpaw\_psl.1.0.0.UPF} available at \url{https://www.quantum-espresso.org/pseudopotentials}. The kinetic energy cutoffs for wavefunctions and for charge density were set to 50 and 300 Rydberg, respectively. All calculations were performed at the $\Gamma$ point. Broyden mixing with local density-dependent Thomas-Fermi screening was utilized, with a mixing factor of 0.05 and 8 mixing iterations. For each nuclei configuration, the energy was computed by a self-consistent field calculation with a threshold of $10^{-4}$ a.u. For ionic dynamics, the `bfgs' algorithm was used. Geometry optimization was carried out using the Davidson algorithm, with the convergence criteria of $10^{-4}$ a.u energy and $10^{-2}$ a.u force. In all cases, the atomic coordinates were allowed to relax while cell parameters were kept fixed.

Isolated molecules (node with 11 and 12 ligands bound, protonated ligand -- H$_2$BDC, H$_2$O and HCl) were treated using Martyna--Tuckerman\cite{martyna1994constant} corrections.

\section{Results}

\subsection{Validation}

To validate the nb-UiO-FF force field, a comprehensive set of tests were performed to assess its accuracy in reproducing structural, thermodynamic and dynamic properties of UiO-66, as described below.

\paragraph{Structural Properties}
Since number density was used as criterion during Morse parameter selection, the force field is already consistent with experimental lattice parameters by construction. The experimental lattice parameter reported for pristine, solvent-free UiO-66 is 20.7004(2) \AA. \cite{Cavka2008,Valenzano2011} For comparison, the present model yields an average lattice parameter of 41.419 $\pm$ 0.017 \AA \ for the 2$\times$2$\times$2 supercell, corresponding to a unit cell value of 20.709 \AA, resulting in a deviation of approximately 0.046\% from experiment and a derived volume deviation of 0.14\%. A comparison with a previously reported model by Rogge \textit{et al.}\cite{Rogge2016}  shows a very close unit cell parameter of 21.117(1) \AA, obtained under slightly different thermodynamic conditions (300 K and \~0.99 atm, versus 298 K and 1 atm in this work). 

\begin{figure}[h!]
    \centering
    \includegraphics[width=\columnwidth]{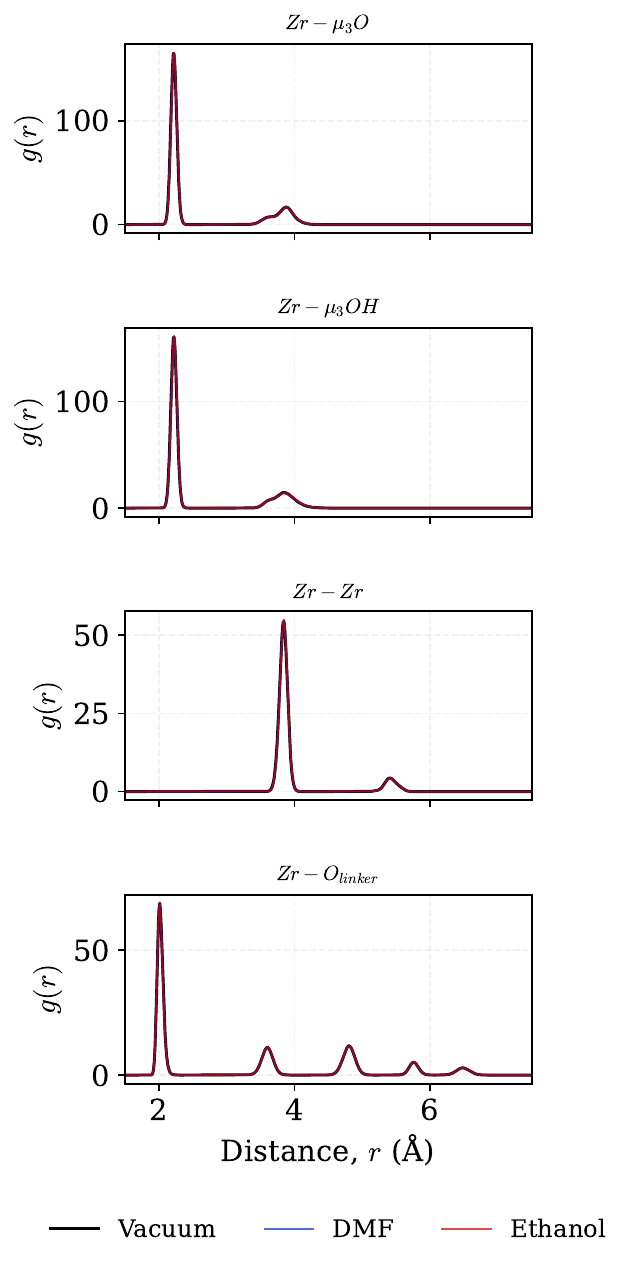}
    \caption{RDFs for selected atomic pairs indicated in each subplot in UiO-66 under vacuum, DMF, and ethanol environments.}
    \label{fig:rdfs}
\end{figure}

To further verify the local structure of the metal cluster and metal-ligand coordination environment, RDFs were computed and compared with reference bond distances obtained from both DFT calculations performed in this work and experimental data,\cite{Valenzano2011} these are shown in Fig. \ref{fig:rdfs}. The positions of the first RDF peaks computed from simulations carried out with nb-UiO-FF show reasonable agreement with previously reported data, with deviations generally within 10$\%$ (see Table \ref{tab:rdf_comparison}). In particular, the RDFs corresponding to  Zr--$\mu_3$O and Zr--$\mu_3$OH exhibit very similar first peak positions, that could only be resolved using a 0.01 \AA \ bin size.

\begin{table}[h!]
    \centering
    \begin{tabular}{c @{\hspace{0.2cm}} c @{\hspace{0.2cm}} c @{\hspace{0.2cm}} c}
    \toprule
    Atom pair & nb-UiO-FF (\AA) & DFT (\AA) & Exp. (\AA) \\
    \midrule
    Zr--$\mu_3$O & 2.215 & 2.06 & 2.118 \\
    Zr--$\mu_3$OH & 2.225 & 2.25 & 2.118 \\
    Zr-Zr & 3.845 & 3.51 & 3.531 \\ 
    Zr--$O_{\mathrm{ligand}}$ & 2.015 & 2.20 & 2.266  \\
    \bottomrule
    \end{tabular}
    \caption{Comparison of RDF peak positions with reference bond distances from DFT (this article)} and experimental data reported in Ref.~\cite{Valenzano2011}.
    \label{tab:rdf_comparison}
\end{table}

\paragraph{Stability in solvents}
The methodology and results presented above were obtained in the absence of solvents, and thus they represent the activated MOF. However, inclusion of solvent is needed to study self-assembly and formation of defects in solvothermal synthesis. Therefore, we tested our models with DMF\cite{Cavka2008, Schaate2011, Zhao2017} and ethanol\cite{Srikantamurthy2025}, solvents that are commonly used in UiO-66 synthesis.

Structural stability was confirmed by the absence of framework distortion over the course of the production trajectory. The lattice parameters were found to be 41.37 \AA \ and 41.40 \AA \ in  DMF and ethanol, respectively. The corresponding RDFs, shown in Fig.~\ref{fig:rdfs}, exhibit nearly complete overlap across vacuum, DMF and ethanol environments, indicating that the structural features of the framework are preserved upon solvation.

\begin{figure}[htbp]
    \centering
    \noindent
    \begin{subfigure}{0.48\textwidth}
        \centering
        \includegraphics[trim=12cm 0 12cm 0, clip, width=\textwidth]{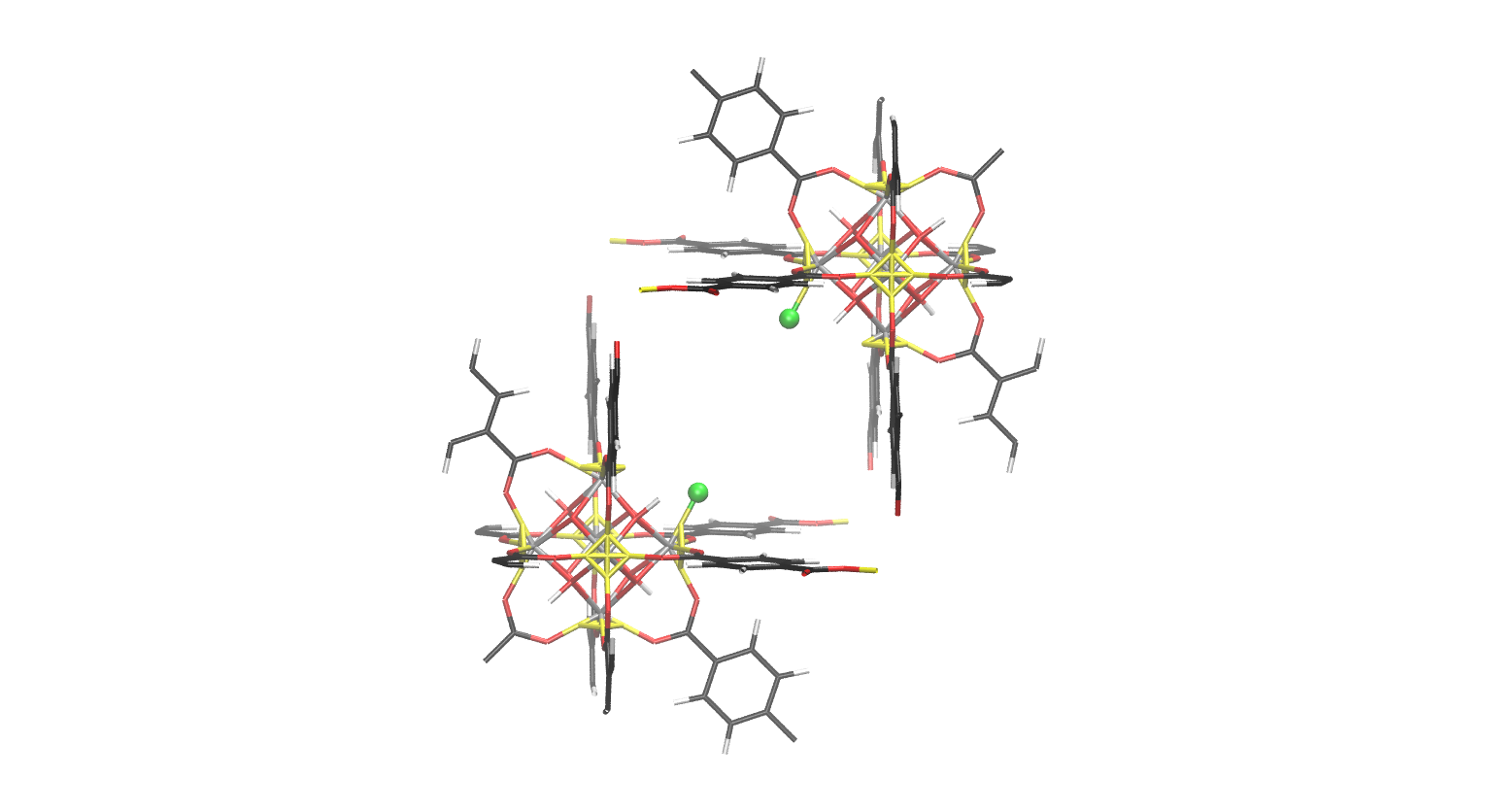}
        \caption{}
        \label{fig:defect_1}
    \end{subfigure}
    \hfill
    \begin{subfigure}{0.48\textwidth}
        \centering
        \includegraphics[trim=12cm 0 10cm 0, clip, width=\textwidth]{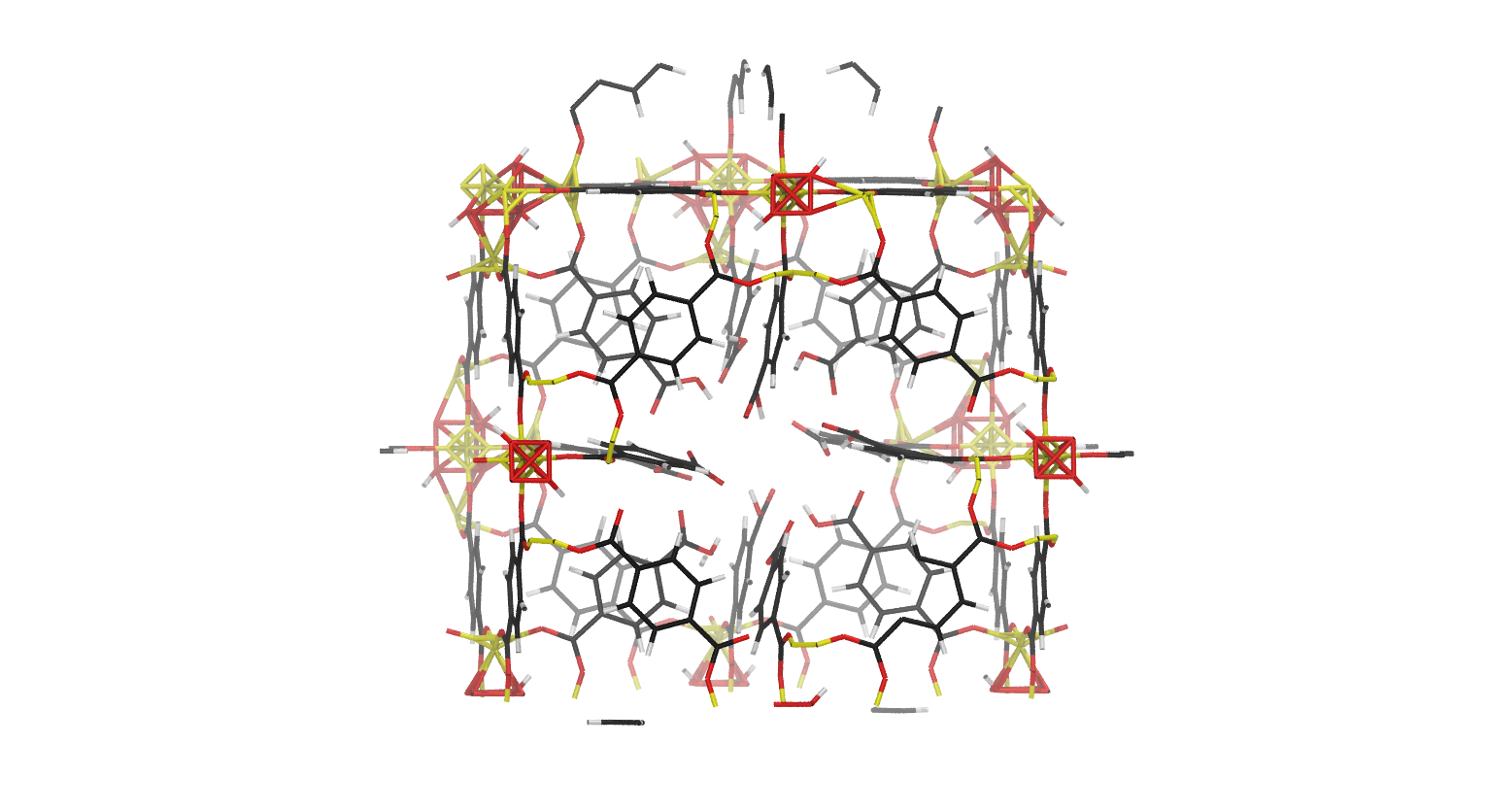}
        \caption{}
        \label{fig:defect_2}
    \end{subfigure}
    
    \caption{Local coordination environment of the two point defects within the framework. (a) Missing Ligand Defect. Chloride ions (green) cap the undercoordinated metal centers. (b) Missing Node Defect. HBDC ligands replace the BDC ligands that would have bound with the missing node.}
    \label{fig:both_defects}
\end{figure}

\paragraph{Stability with defects}
Point defects are an important feature of UiO-66, and it is therefore essential that the developed force field is able to accurately describe defective structures. Two defect types were considered: missing-linker and missing-node, introduced by removing the corresponding structural units from a 2$\times$2$\times$2 supercell of the pristine framework (see Fig.~\ref{fig:both_defects}). Charge neutrality was preserved by capping the undercoordinated metal centers with chloride ions ($Cl^-$) in the case of missing-linker defects, following previous studies \cite{Bristow2014} that indicated that $Cl^-/H_2O$ charge-capping mechanisms are energetically favored. Monodentate benzene-1,4-dicarboxylate (HBDC) fragments were introduced in the case of missing-node defects. The net charge of the removed species was redistributed among the capping atoms to ensure that the total charge of the system remains unchanged. All simulations were carried out including DMF or ethanol molecules as solvents. These simulations revealed that our force field is able to model stable missing-ligand or missing-node frameworks in both solvents for at least 20 ns.

To perform a quantitative comparison with experimental data,\cite{Oien2014}  the relative change in lattice parameter upon defect introduction was evaluated. In experimental studies at high defect concentrations (73$\%$ linker occupancy), the cell length increases from 20.7004(2) \AA \ to 20.7465(2) \AA, corresponding to a relative expansion of approximately $0.22\%$. In our simulations, we have one single missing linker in a 2$\times$2$\times$2 supercell ($\sim0.5$ \% defect concentration, much lower than the experimental one of $\sim25$\%) which resulted in a cell length of 41.422 $\pm$ 0.017 \AA. The difference between the pristine and missing linker (~0.003 \AA) is significantly smaller than the intrinsic fluctuations of the cell length. Thus, higher defect concentrations were investigated, to see whether the cell expansion that was experimentally observed could be reproduced. To this end, a 25\% defect concentration system was studied. The MOF was able to sustain this high defect concentration and remained stable throughout our simulations. Moreover, we found that the cell length strongly depends on the relative position of the defects. Thus, we modelled 3 systems with random orientation of defects. Over the 3 systems, the cell lengths were averaged and found to be 41.4200 $\pm$ 0.0214 \AA, with two systems exhibiting a clear cell expansion and the third one showing a contraction. Therefore, to derive a statistically significant trend for cell expansion, all compatible microstates for this system should be considered, which is out of the scope of this work.

\paragraph{Energetics validation}

To validate the energetics, both the coordination bond and the defect generation energies were computed using nb-UiO-FF and compared to those obtained via DFT calculations for a set of comparable chemical reactions. These comparisons should be considered as semi-quantitative, as the chemical reactions are not exactly the same in both cases for reasons that will be detailed below.

To calculate the coordination bond energy, the energy of the following chemical reactions were computed. For DFT calculations, in the absence of one ligand, the vacant coordination site on the node is capped by an $OH^-$ ion to maintain charge neutrality.

\begin{equation}
\begin{aligned}
&\mathrm{Zr}_6(\mu_3\text{-O})_4(\mu_3\text{-OH})_4
(\mathrm{HBDC})_{11}\mathrm{OH}
+ \mathrm{H_2BDC} \\
&\qquad\rightarrow
\mathrm{Zr}_6(\mu_3\text{-O})_4(\mu_3\text{-OH})_4
(\mathrm{HBDC})_{12}
+ \mathrm{H}_2\mathrm{O}
\end{aligned}
\end{equation}

 For nb-UiO-66, however, the OH$^-$ capping was not included in the force-field model because the newly introduced metal--OH$^-$ bond was not parameterized. Thus, the following reaction was considered to compare with the DFT results:

\begin{equation}
\begin{aligned}
&\mathrm{Zr}_6(\mu_3\text{-O})_4(\mu_3\text{-OH})_4
(\mathrm{HBDC})_{11}^+
+ \mathrm{HBDC^-} \\
&\qquad\rightarrow
 \mathrm{Zr}_6(\mu_3\text{-O})_4(\mu_3\text{-OH})_4
 (\mathrm{HBDC})_{12}
 \end{aligned}
 \end{equation}

The DFT resulted in -23.22 kcal/mol energy, while nb-UiO-FF yielded a comparable energy of -32.33 kcal/mol. Therefore, the node--ligand binding energy from nb-UiO-FF is in reasonable agreement to DFT values.

Defect formation energies were studied by comparing the energies of the unit cell of UiO-66 and a defective one with a missing linker defect, and capping the corresponding nodes with $Cl^-$ in the case of DFT. The following chemical reaction was considered:

\begin{equation}
\begin{aligned}
&\left[\mathrm{Zr}_6(\mu_3\text{-O})_4(\mu_3\text{-OH})_4\right]_4
(\mathrm{BDC})_{24}
+ 2\mathrm{HCl} \\
&\qquad\rightarrow
\left[\mathrm{Zr}_6(\mu_3\text{-O})_4(\mu_3\text{-OH})_4\right]_4
(\mathrm{BDC})_{23}(\mathrm{Cl})_2 \\
&\qquad\quad + \mathrm{H}_2\mathrm{BDC}
\end{aligned}
\end{equation}

It gave an energy of 33.5 kcal/mol. To compare with nb-UiO-FF, a unit cell comparison was not feasible because of the short range cutoff interaction, which is larger than half the unit cell length. Thus, we compared with a $2\times2\times2$ supercell with 8 ligands removed, to keep the same defect concentration as in the DFT calculation. Three different random defect orientations were modeled and their potential energy was averaged for statistical significance. The following reaction energy was calculated:

\begin{equation}
\begin{aligned}
&\left[\mathrm{Zr}_6(\mu_3\text{-O})_4(\mu_3\text{-OH})_4\right]_{32}
(\mathrm{BDC})_{192}
+ 16\mathrm{HCl} \\
&\qquad\rightarrow
\left[\mathrm{Zr}_6(\mu_3\text{-O})_4(\mu_3\text{-OH})_4\right]_{32}
(\mathrm{BDC})_{184}(\mathrm{Cl})_{16} \\
&\qquad\quad + 8\mathrm{H}_2\mathrm{BDC}
\end{aligned}
\end{equation}

This process requires an energy of 154.54 kcal/mol, and the per defect creation energy is 19.32 kcal/mol, which is close to the DFT value.

We also tested modelling one missing linker defect in the $2\times2\times2$ supercell, which follows the process:

\begin{equation}
\begin{aligned}
&\left[\mathrm{Zr}_6(\mu_3\text{-O})_4(\mu_3\text{-OH})_4\right]_{32}
(\mathrm{BDC})_{192}
+ 2\mathrm{HCl} \\
&\qquad\rightarrow
\left[\mathrm{Zr}_6(\mu_3\text{-O})_4(\mu_3\text{-OH})_4\right]_{32}
(\mathrm{BDC})_{191}(\mathrm{Cl})_{2} \\
&\qquad\quad + \mathrm{H}_2\mathrm{BDC}
\end{aligned}
\end{equation}

The process has an energy of 26.27 kcal/mol. This defect formation energy is also comparable with our DFT calculation. In addition, these results show that defect interactions are not negligible, which is reasonable given the close proximity between defects in high defect concentrations as those modeled in this study.

\paragraph{Mechanical Properties}
To evaluate the ability of nb-UiO-FF in estimating the mechanical properties of UiO-66, the elastic constants were computed using the energy--strain method\cite{Kaur2023StressStrain} by applying small, finite deformations to the simulation cell and monitoring the corresponding change in potential energy. For the cubic UiO-66 structure there are only three independent non-zero elastic constants ($C_{11}$, $C_{12}$, and $C_{44}$), as shown in Equation \ref{eq:mech}:

\begin{equation}
\begin{split}
E &= \tfrac{1}{2} C_{11} \left( e_{xx}^2 + e_{yy}^2 + e_{zz}^2 \right) \\
  &\quad + \tfrac{1}{2} C_{44} \left( e_{xy}^2 + e_{yz}^2 + e_{zx}^2 \right) \\
  &\quad + C_{12} \left( e_{xx} e_{yy} + e_{yy} e_{zz} + e_{xx} e_{zz} \right)
\end{split}
\label{eq:mech}
\end{equation}

Three types of deformation: uniaxial, biaxial, and shear strains were applied to isolate these elastic constants. Uniaxial strain was applied to determine $C_{11}$, shear deformation was used to extract $C_{44}$, and biaxial strain, in combination with the previously obtained $C_{11}$, was employed to evaluate $C_{12}$. 

The resulting elastic constants are higher than the DFT values from the work by Wu \textit{et al.}\cite{Wu2013} (see Tab.~\ref{tab:elast-const}) especially $C_{44}$. This could be attributed to the presence of strong bonded interactions between the Zr center and the dummy atoms in the CDA model, which effectively increase the rigidity of the metal-ligand coordination environment.

\begin{table}[h]
    \centering
    \begin{tabular}{c @{\hspace{0.5cm}} c @{\hspace{0.5cm}} c @{\hspace{0.5cm}} c}
    \toprule
    Elastic Constants & DFT (GPa) & nb-UiO-FF\\
    \midrule
    $C_{11}$ & 59.35 & 63 \\
    $C_{12}$ & 31.85 & 34 \\
    $C_{44}$ & 17.63 & 24 \\
    \bottomrule
    \end{tabular}
    \caption{Comparison of elastic constants with DFT benchmarks from Ref. \citenum{Wu2013}.}
    \label{tab:elast-const}
\end{table}

The elastic constants were thereafter used to derive the macroscopic mechanical properties of the material, which are reported in Tab. S11.  The bulk modulus obtained was 44 GPa, compared to 41.0 GPa from DFT \cite{Wu2013}. This value is $\sim$2.5 times higher than the experimentally reported value of 17.0(1.5) GPa, \cite{Yot2015} which corresponds to an 11-fold coordinated (defective) UiO-66 sample. It should be noted that no experimental bulk modulus has been reported for non-defective UiO-66, limiting direct comparison. Overall, nb-UiO-FF overestimates the rigidity of the MOF and should thus be used with caution for evaluating mechanical properties.

\paragraph{Transferability to isoreticular MOFs}

\begin{figure}
    \centering
    \includegraphics[trim=250px 60px 250px 60px, clip, width=\linewidth, keepaspectratio]{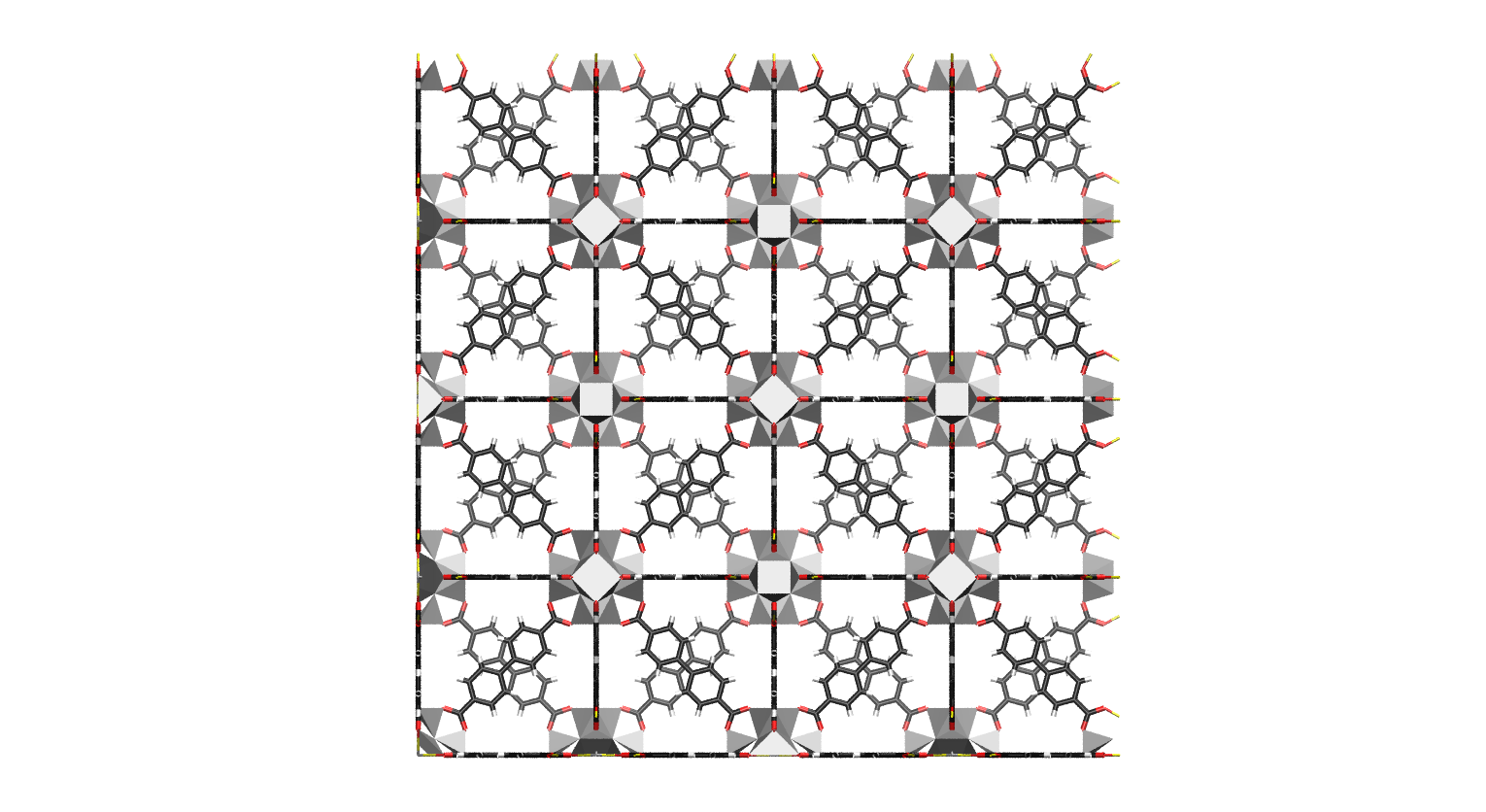}
    \caption{2$\times$2$\times$2 supercell of UiO-67.}
    \label{fig:UiO-67}
\end{figure}

We further evaluated the transferability of the developed force field for an isoreticular MOF: UiO-67\cite{Cavka2008} (see Fig. \ref{fig:UiO-67}). Isoreticular MOFs share the same underlying topology and metal cluster connectivity but differ in the nature of the organic linkers, enabling systematic variation of pore size. UiO-67 belongs to the same isoreticular family than UiO-66, comprising  biphenyl-4,4$^{\prime}$-dicarboxylate (BPDC) linkers and the same metal cluster as UiO-66. Simulations confirmed that the framework remains structurally stable. The cell parameter was averaged over the final 15 ns of the production run, and was found to be 53.28 $\pm$ 0.02 \AA. The corresponding experimental cell parameter for the 2 × 2 × 2 supercell is 53.7618 \AA,\cite{Oien2014} showing a deviation of 0.9\% in the cell length for the force field predicted value. The RDFs were also calculated and are shown in Fig. S7. The RDF peak positions for Zr--$\mu_3$O, Zr--$\mu_3$OH, and Zr--$O_{carboxyl}$ were found to be 2.235, 2.255, and 2.015 \AA, respectively. These values are in reasonable agreement with the previously reported bond lengths of 2.058, 2.251, and 2.219 \AA, respectively,\cite{Oien2014} with deviations of up to approximately $9\%$. In addition, we also reported $Zr-Zr$ and $Zr-C_{carboxyl}$ RDFs. The same procedure that we applied in this section for UiO-67 could be applied to model UiO-68, another member of the isoreticular family of UiO-66.  

\subsection{Self-assembly studies}

Once the force field validated, we examined the early steps of the self-assembly of UiO-66, to uncover the building units that are formed via the node--ligand binding events that happen at the onset of the process. Simulations were performed from an initial configuration built by randomly inserting 16 \ce{Zr} oxoclusters, 96 BDC ligands and 2136 DMF solvent molecules (to reproduce the bulk density of DMF at 298 K and 1 atm), in a box of length 65 \AA , which corresponds to concentrations of 0.097 M and 0.580 M for the metal clusters and ligands respectively, and a total of 15024 atoms. To ensure statistical robustness, three independent simulations were carried out using distinct initial configurations. 

\begin{figure}[htbp] 
    \centering

    \begin{subfigure}[t]{0.32\textwidth}
        \centering
        \includegraphics[trim=370px 40px 370px 40px, clip, width=\linewidth]{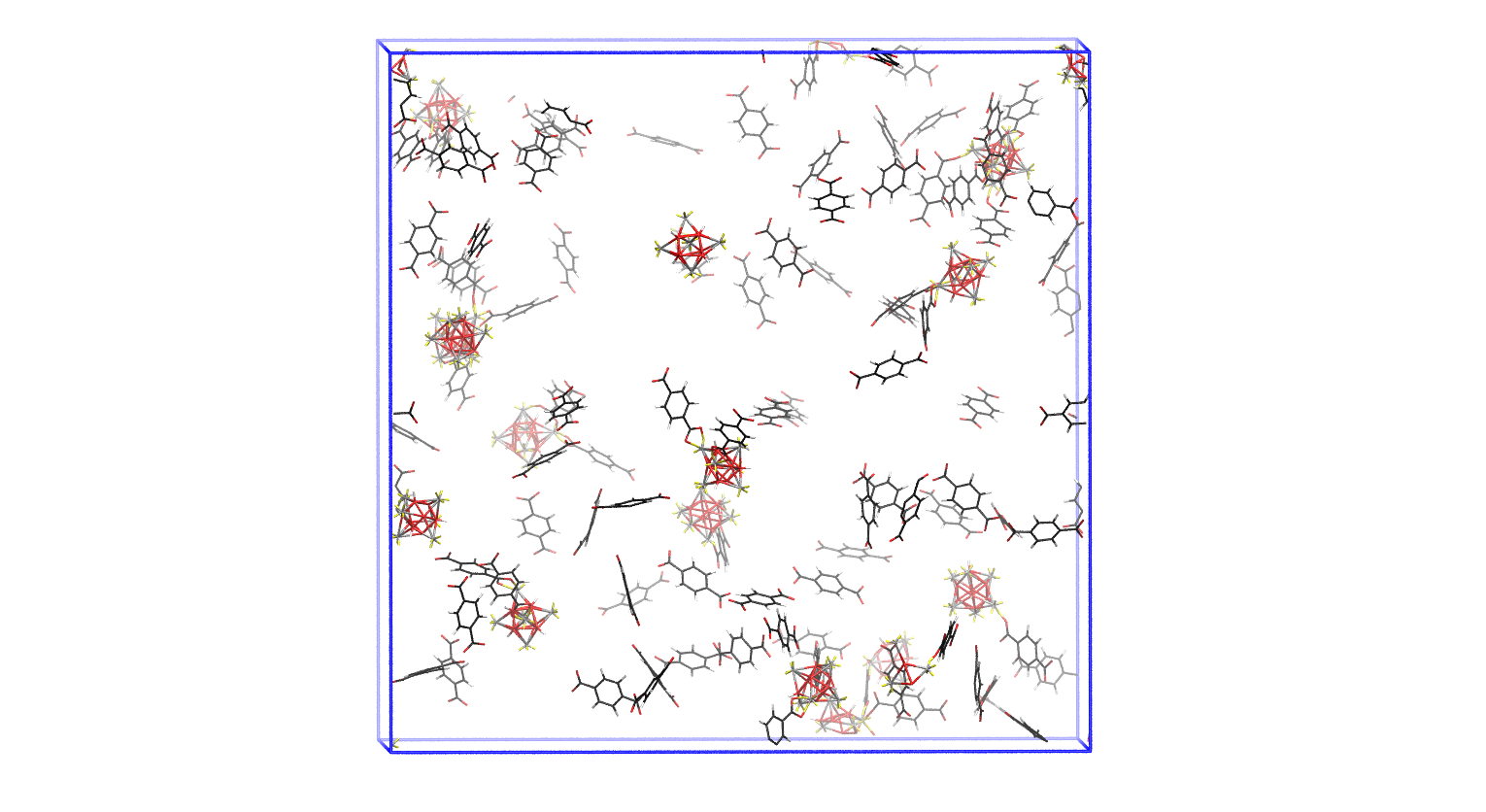}
        \caption{}
    \end{subfigure}\hfill
    \begin{subfigure}[t]{0.32\textwidth}
        \centering
        \includegraphics[trim=370px 40px 370px 40px, clip, width=\linewidth]{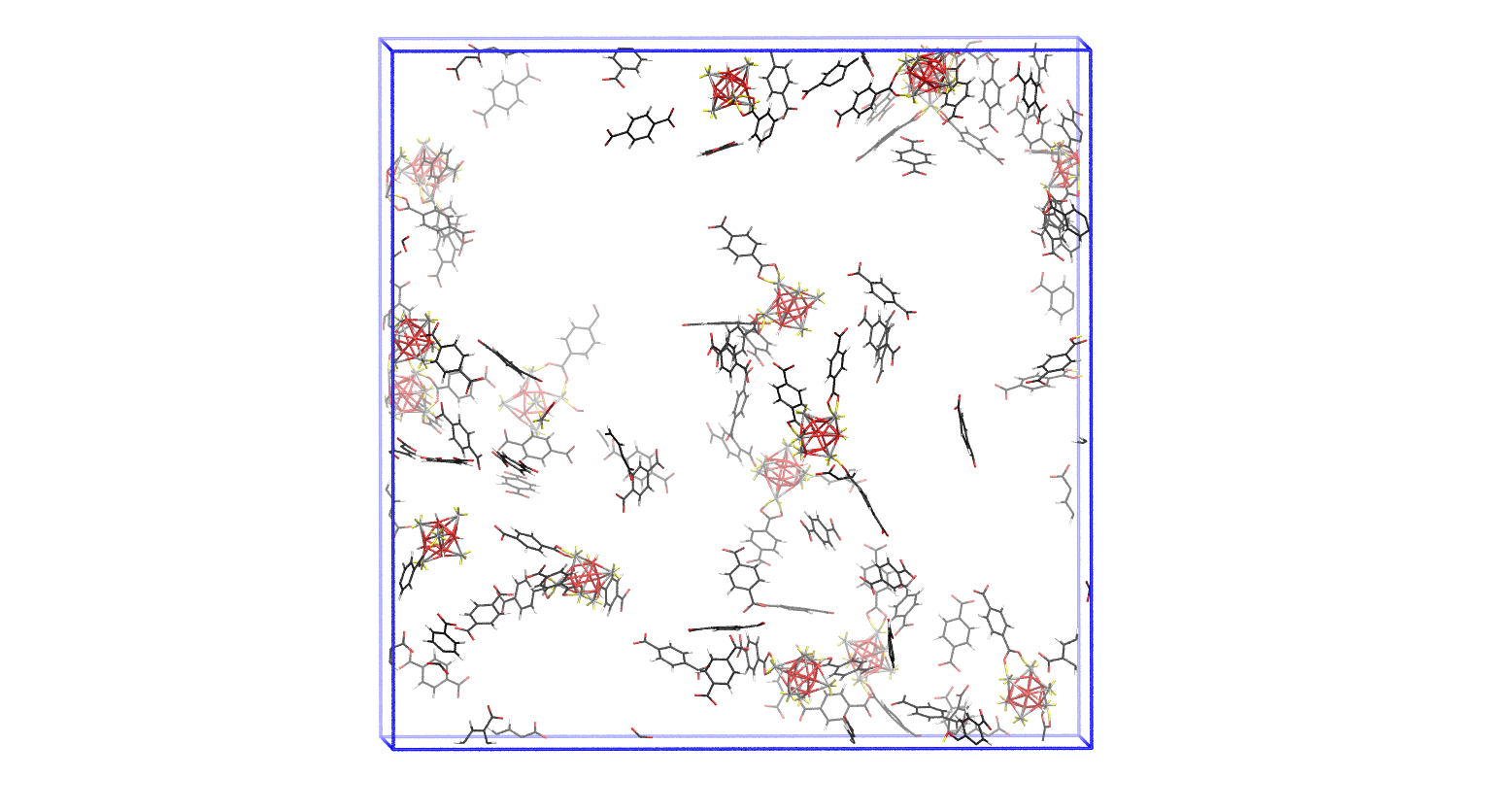}
        \caption{}
    \end{subfigure}\hfill
    \begin{subfigure}[t]{0.32\textwidth}
        \centering
        \includegraphics[trim=370px 40px 370px 40px, clip, width=\linewidth]{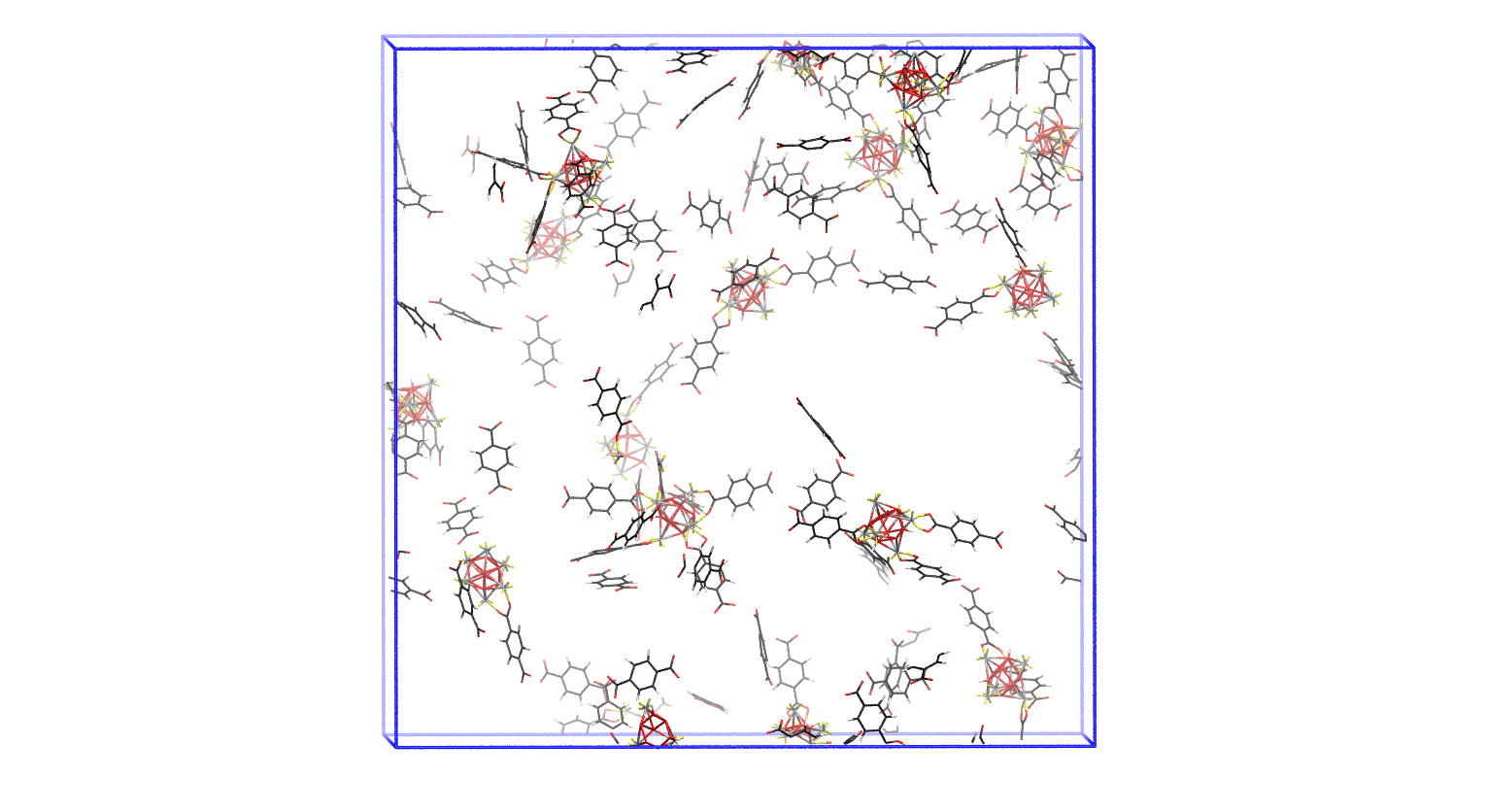}
        \caption{}
    \end{subfigure}

    \caption{Snapshots of UiO-66 self-assembly. Solvent was omitted for clarity. (a) At the starting of the simulations, (b) 3 ns and (c) 6 ns.}
    \label{fig:trajectory}
\end{figure}

Snapshots illustrating the evolution of the early stages of the self-assembly process in DMF are shown in Fig \ref{fig:trajectory}. The solvent is ommited for clarity. The formation of coordination bonds between metal clusters and organic linkers can be clearly observed over the course of the trajectory. By comparing the configurations at t = 0 ns and t = 6 ns, it can be seen that the amount of free ligands is considerably reduced over time (panel a versus panel c in Fig. \ref{fig:trajectory}). 

To obtain a quantitative measure of the degree of progress of the formation reaction, the  coordination number between \ce{Zr} and linker oxygen atoms was monitored and is given in Fig. S8. The coordination number increases over time for the three independent trajectories, indicating that node--ligand bonds are progressively formed over time. After 6 ns, all trajectories show average Zr--O$_{ligand}$ coordinations in the range of $\sim$0.55-0.7, which compared with the value of 4 that is attained in bulk UiO-66 is very low. This shows that these simulations only sample the very early stages of self-assembly. Still, it should be noted that the coordination number would not be expected to reach 4 in these simulations, even if they were considerably longer. This is because the amount of solvent that is present inside the pores of a 2x2x1 UiO-66 supercell is lower than the number of solvent molecules in this simulation box (24 molecules per unit cell -- see Fig.~S6, so a total of 96 molecules, versus 2136). Since our simulations are performed at constant number of molecules, the synthesis process will happen under an excess of solvent which cannot be displaced as it would in a real synthesis experiment, due to limitations in the simulation box size. Thus, the maximum that we could aspire to get within this simulation scheme would be a highly-connected cluster, including most of the nodes and ligands, as we observed previously for ZIF-8 in similar setups.\cite{Balestra2022,AndarziGargari2025} This highly-connected cluster will have a surface with Zr ions that are undercoordinated, and thus, Zr--O$_{ligand}$ average coordination will never reach the bulk value of 4 in these simulations. Moreover, aggregation of the metal clusters is only observed once within these simulations, which can be attributed to their comparatively slow diffusion arising from their larger mass and size. This prevents them from reaching inter-node binding positions within the short simulation timescales. Despite these limitations, these simulations allow to sample some of the early species that are formed in the UiO-66 synthesis process, which was inaccessible with previous force fields and is extremely difficult to measure experimentally.
 
\begin{figure}[htbp] 
    \centering

    \begin{subfigure}[t]{0.48\textwidth}
        \centering
        \includegraphics[trim=380px 70px 380px 70px, clip, width=0.7\linewidth]{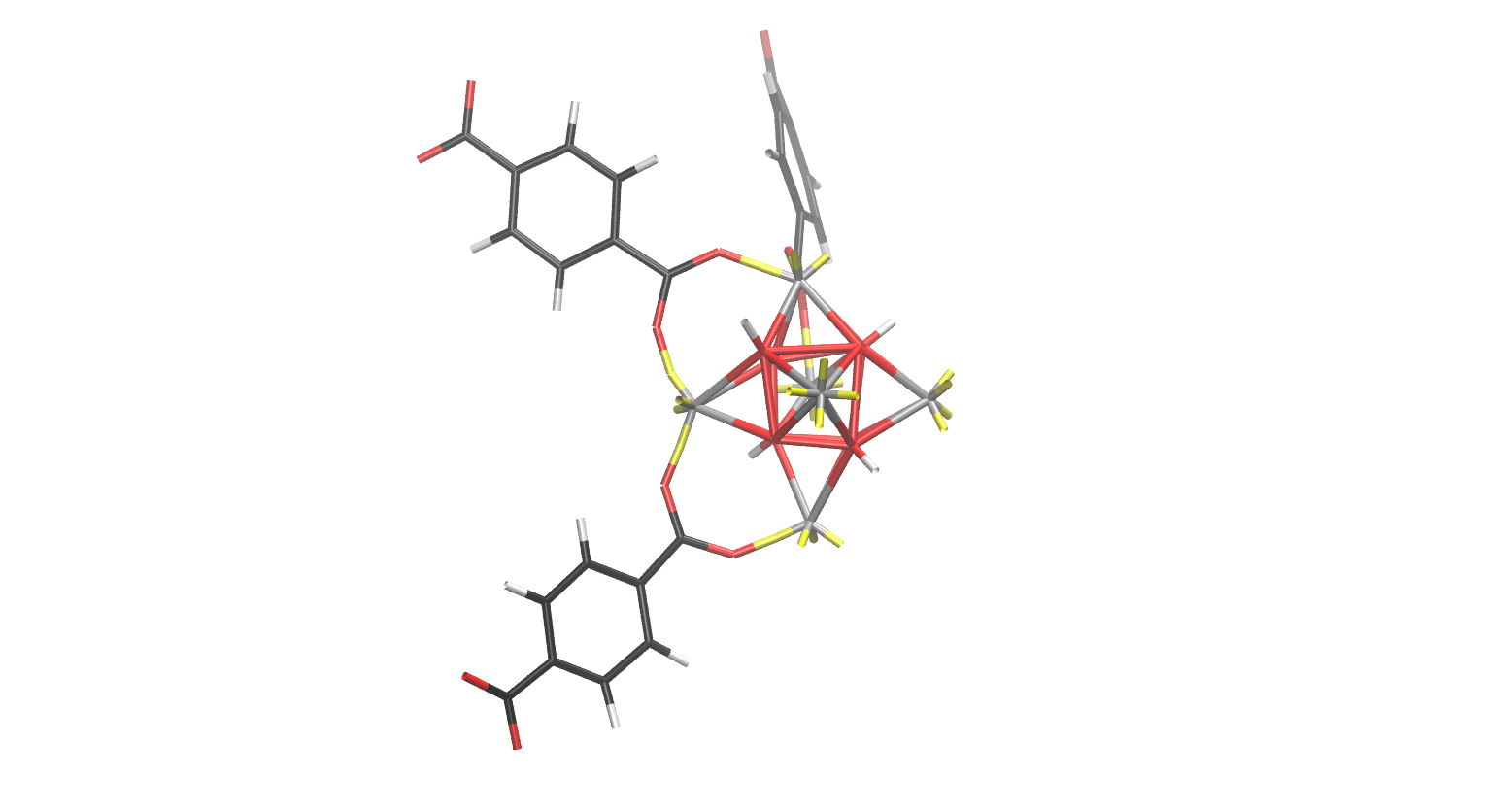}
        \caption{}
        \label{fig:SBU_motif}
    \end{subfigure}\hfill
    \vspace{10px}
    \begin{subfigure}[t]{0.48\textwidth}
        \centering
        \includegraphics[trim=400px 50px 400px 30px, clip, width=0.7\linewidth]{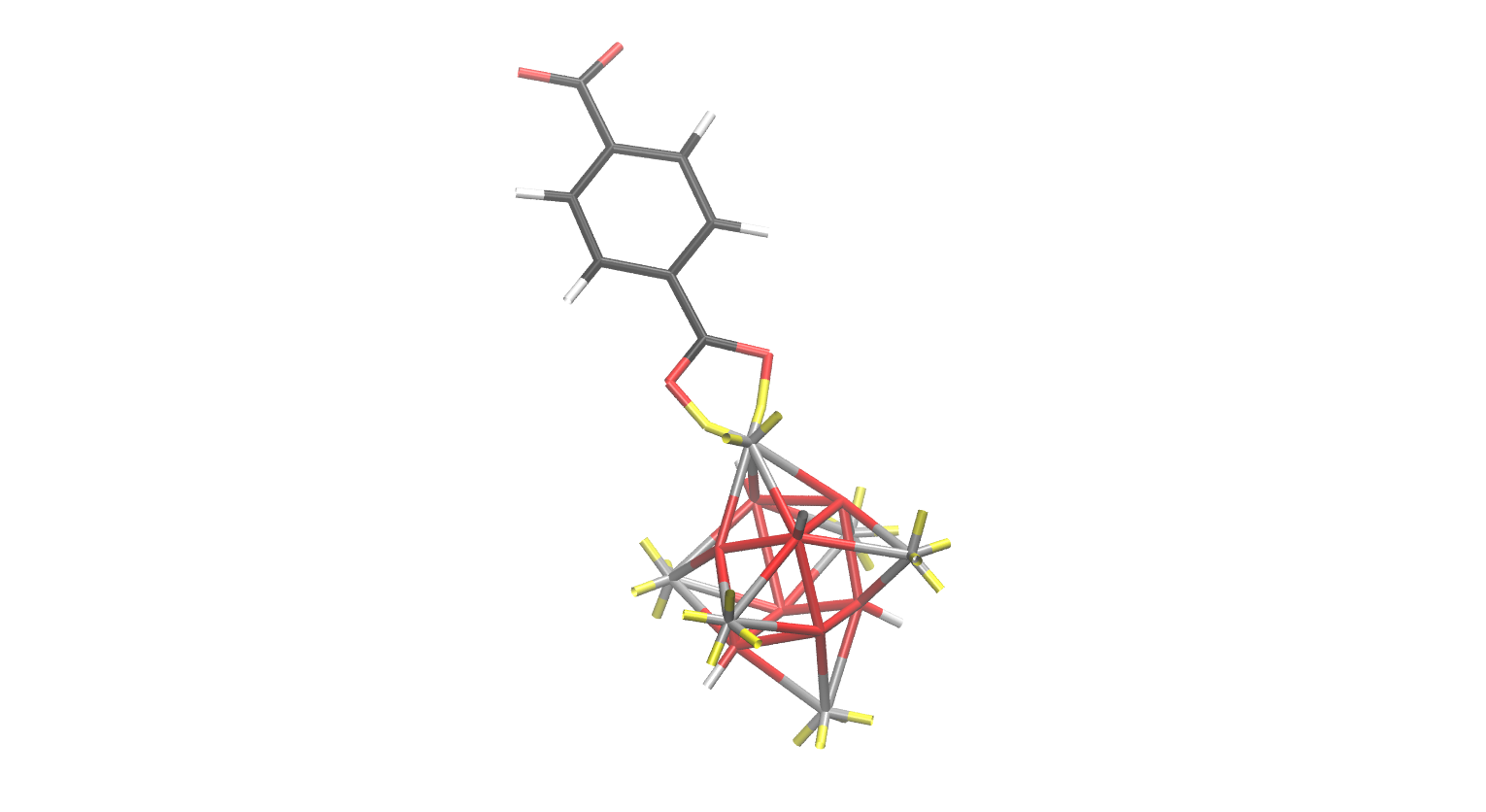}
        \caption{}
        \label{fig:binding}
    \end{subfigure}

    \caption{Selected representative structural motifs observed during the simulations. (a) SBU motif. (b) Metastable \ce{Zr}--ligand binding configuration.} 
    \label{fig:motifs}
\end{figure}

Examples of two motifs that were found in the trajectories are provided in Fig. \ref{fig:motifs}. Fig. \ref{fig:SBU_motif} shows a typical secondary building unit (SBU) of UiO-66. This early formed species, formed by a node bound to three ligands, presents a structural pattern that is present in the bulk MOF that forms at the end of the synthesis process. In addition, non-ideal binding configurations are also observed, where both oxygen atoms on the same carboxyl group of a BDC linker interact with two dummy atoms of the same \ce{Zr} atom as shown in Fig. \ref{fig:binding}. This configuration is not present in the UiO-66 crystal; however, a similar binding configuration has been reported by Semivrazhskaya and coworkers \cite{Semivrazhskaya2023} for a family of zirconium chloroterephthalates with the general formula [ZrCl$_{5-x}$(HBDC)$_x$]$^{-}$, where $1 < x < 5$. These species form upon dissolution of zirconium chloride and terephthalic acid, with partial substitution of chloride ligands by terephthalate. Although the exact species differ from the ones found in our simulation, their existence suggests that chelation involving both oxygen atoms of a single carboxyl group is chemically favorable.
These species likely correspond to transient or kinetically trapped states within the self-assembly process. 

\begin{figure}
    \centering
    \includegraphics[width=\linewidth]{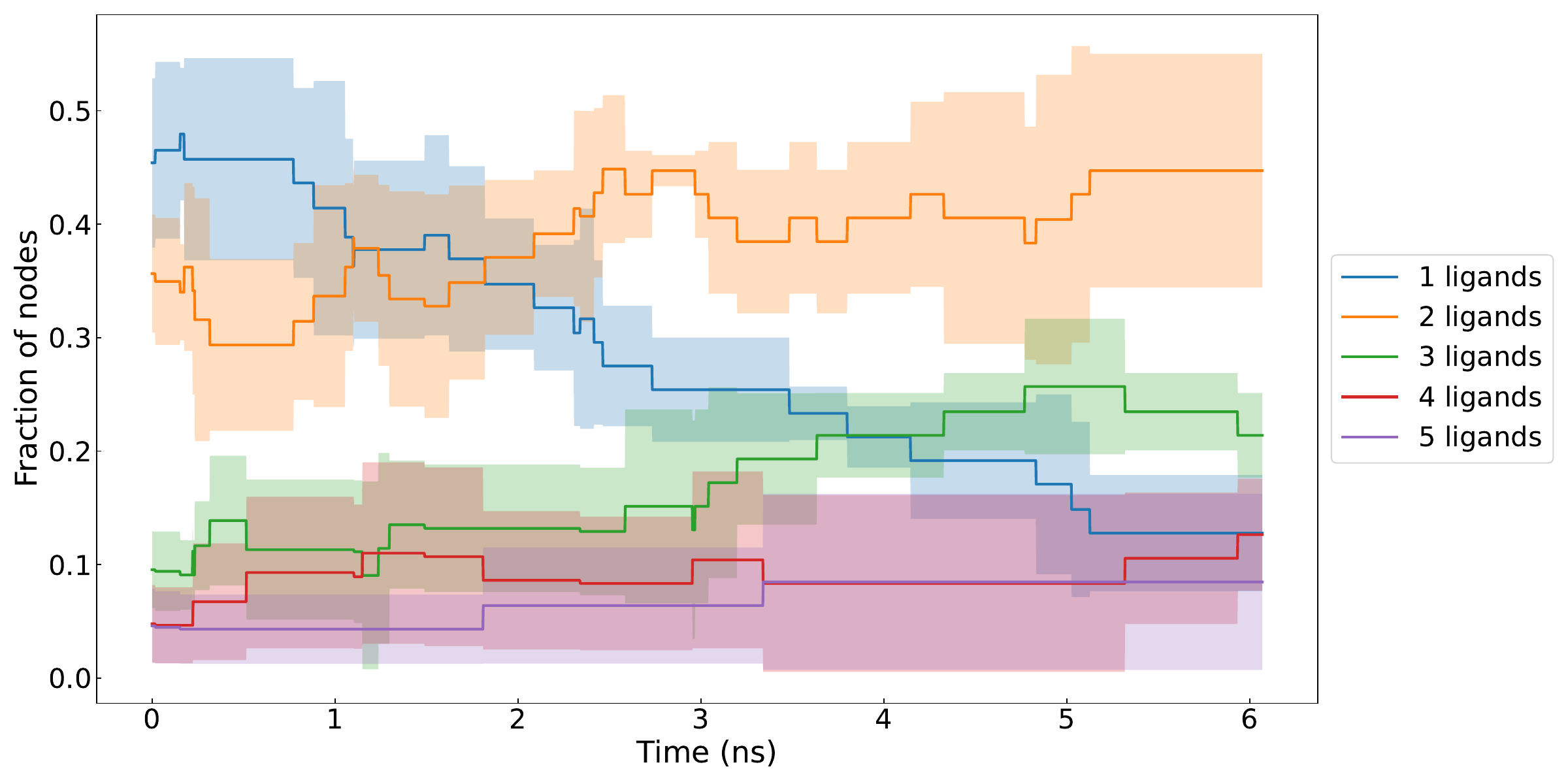}
    \caption{Time evolution of the fraction of nodes with different coordination numbers (1–5 ligands). Solid lines represent the mean over three independent trajectories, while shaded regions indicate the corresponding standard deviation.}
    \label{fig:species_pop}
\end{figure}

To study the whole range of species formed during the early stages of UiO-66 self-assembly, the evolution of the fraction of nodes coordinated by a given number of ligands was monitored, and its average for the three independent simulation trajectories is shown in Fig. \ref{fig:species_pop}. Secondary building units ranging from 1- to 5- connected nodes are observed over the course of the simulation, showcasing the variety of intermediate species that can be formed at the early stages of UiO-66 self-assembly. The number of nodes that are bound to a single ligand (blue curve) decreases along the simulation, reaching values of $\sim$11\% by t = 6 ns. This decrease is compensated by the formation of additional node--ligand bonds. $\sim$45\% and $\sim$20\% of the nodes are connected to 2- and 3- ligands respectively, being the former the most abundant species by the end of the simulations in all cases. The maximum node coordination observed in these simulations is five ligands, which is lower than the ideal coordination number of twelve expected for a perfectly coordinated structure. This reduced coordination is due to the fact that only the onset of the self-assembly process is sampled in these trajectories. The dynamics of node--ligand binding do not seem to support a sequential binding process: indeed, 2- and 3-coordinated nodes seem to increase with time in a parallel fashion. This is in contrast to what was observed for the ZIF-8 MOF in our previous works, in which the formation of 3-coordinated metal ions was generally at the expense of a reduction in 2-coordinated ion concentration.\cite{Balestra2022,AndarziGargari2025}

To study further self-assembly stages, it is necessary to overcome kinetic traps and to enable the formation of ionic aggregates containing more than one node. To achieve this, either coarse graining methods\cite{Mondal2026} or enhanced sampling simulations\cite{Laio2002,Barducci2008} are needed. The former acts reducing the degrees of freedom in the simulations, while the latter allow to overcome free energy barriers by adding an on-the-fly dynamical bias to prevent the system to stay trapped in configurations that were previously explored. In future work, we will combine nb-UiO-FF with these techniques to develop an integral framework to study further stages of UiO-66 nucleation.

\section{Conclusions} 
In this work, a transferable, partially reactive force field was developed and validated for the simulation of the UiO-66 and UiO-67 MOFs. The novelty of the force field is that it incorporates node--ligand reactivity, thus enabling the study of self-assembly and defect formation processes via molecular simulation. The parameterization was performed to satisfy a combination of bulk thermodynamic  and structural properties, energetics from DFT calculations and structure of local coordination environments. The resulting model accurately reproduces the experimental lattice parameters of pristine UiO-66 by construction, with deviations within less than $1\%$, and captures the metal-ligand coordination environments as confirmed by the RDFs. Furthermore, the force field demonstrates robust performance in the presence of defects, both missing-linker and missing-node configurations and in two different solvents (ethanol and DMF), both in terms of structure and energetics. Transferability was further assessed on UiO-67 where the correct framework structure was reproduced. A limitation of the force field is that it predicts an overly rigid framework, when comparing elastic constants to DFT values. Once validated, the force field was used to explore the early stages of the UiO-66 self-assembly process, where progressive coordination between metal clusters and organic linkers was observed, highlighting its capability to describe dynamic bond formation. Secondary building units ranging from 1- to 5- connected nodes are described, and kinetically trapped motifs involving the binding of both O in a BDC carboxylate group with a single Zr atom are identified. Overall, the developed force field provides a consistent and physically reliable description of these MOFs and offers a solid foundation for studying its self-assembly behaviour and defect formation.

\section*{Conflicts of interest}
There are no conflicts of interest to declare.

\section*{Data availability}

The data supporting this article have been included as part of the Supplementary Material.

\begin{acknowledgments}
This work was funded by the European Union ERC Starting grant MAGNIFY, grant number 101042514. This work was granted access to the HPC resources of IDRIS under the allocations and A0170915688 and A0190915688 made by GENCI.
\end{acknowledgments}

\bibliography{magnify}

\clearpage
\onecolumngrid

\begin{center}
    \textbf{\huge Supporting Information} \\
    \vspace{6mm}
    \textbf{\huge \textbf{Partially reactive force field for the UiO-66 metal-organic framework}} \\
    \vspace{8 mm}
    {\Large Akanksha Nawani, Rocio Semino} \\
    \vspace{4 mm}
    \large \textit{Sorbonne Université, CNRS, Physico-chimie des Electrolytes et Nanosystèmes Interfaciaux, PHENIX, F-75005 Paris, France } \\
    \vspace{5 mm}
    
\end{center}

\setcounter{section}{0}
\setcounter{figure}{0}
\setcounter{table}{0}
\setcounter{equation}{0}
\renewcommand{\thesection}{S\arabic{section}}
\renewcommand{\thefigure}{S\arabic{figure}}
\renewcommand{\thetable}{S\arabic{table}}
\renewcommand{\theequation}{S\arabic{equation}}

\vspace{5mm}

\section{Force Field Parameters and Schematics}

\begin{figure}[htbp]
    \centering
    \includegraphics[width=0.8\linewidth]{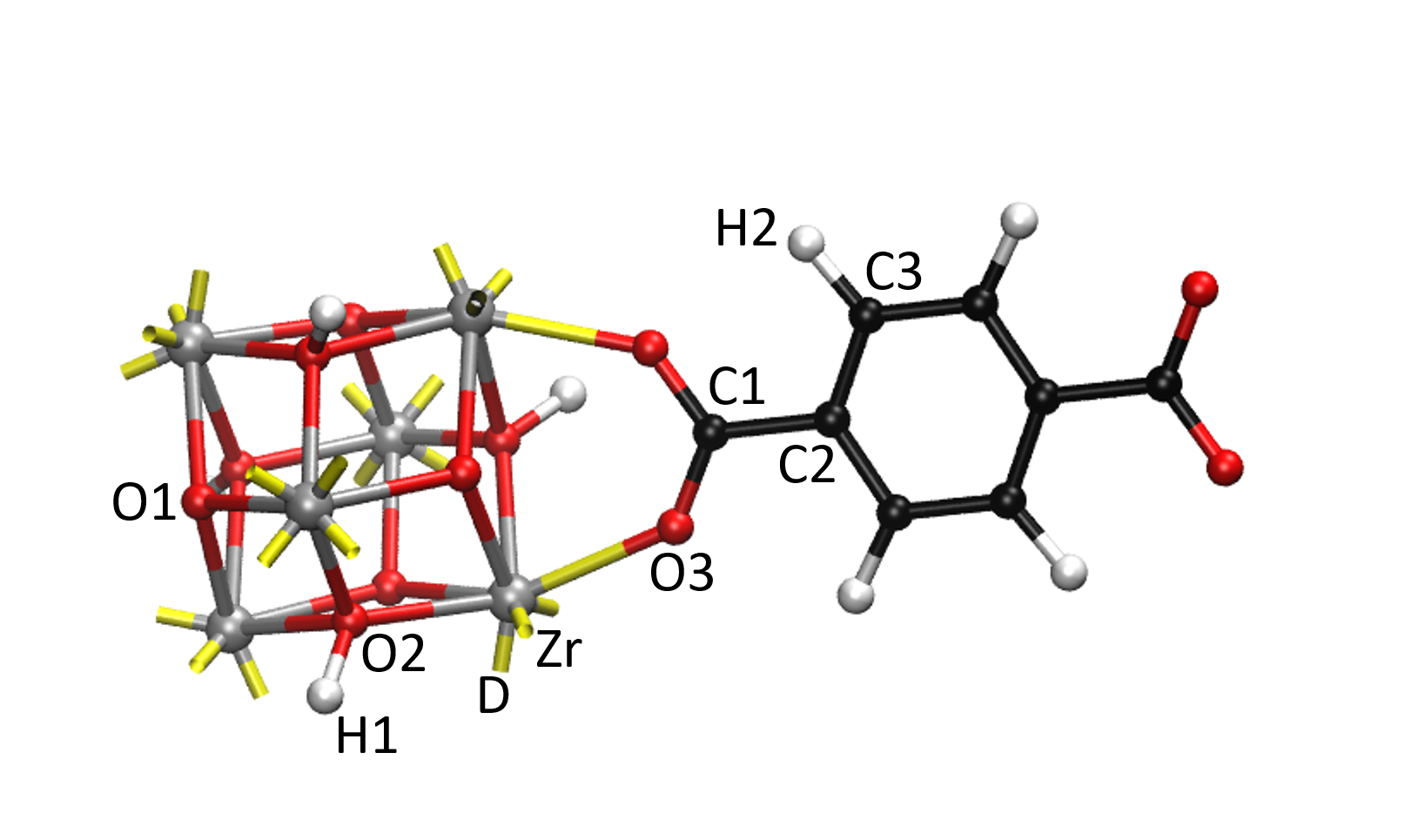}
    \caption{Definition of atom types used in the force field for UiO-66. Representative atoms are labeled according to their chemical environment. Color code: O (red), H (white), Zr (silver), C (black), D (yellow).}
    \label{fig:atom_types}
\end{figure}

Figure S1: this figure illustrates the position of the different atom types in the node and ligand structures, these atom types will be used in the force field tables.

\FloatBarrier

\begin{table*}[htbp]
\centering
\caption{Bond stretching and angle bending parameters for UiO-66.}
\label{tab:bond_angles} 
\begin{tabularx}{\textwidth}{X X X}
\toprule
\midrule

\textbf{Bonds} & \textbf{$K$ [kcal/(mol$\cdot$\AA$^2$)]} & \textbf{$r_0$ [\AA]} \\
\midrule
Zr--O1  & 128.745 & 2.089 \\
Zr--O2  & 128.745 & 2.286 \\
O2--H1 & 540.634 & 1.400 \\
Zr--D & 800     & 0.9   \\
C1--O3 & 540     & 1.273 \\
C1--C2 & 351.253 & 1.487 \\
C2--C3 & 480     & 1.393 \\
C3--C3 & 480     & 1.393 \\
C3--H2 & 363.416 & 1.08  \\

\midrule

\textbf{Angles} & \textbf{$K$ [kcal/(mol$\cdot$rad$^2$)]} & \textbf{$\theta_0$ [deg]} \\
\midrule
O1--Zr--O1 & 13.836 & 102.40 \\
O1--Zr--O2 & 13.836 & 66.881 \\
O2--Zr--O2 & 13.836 & 102.40 \\
Zr--O1--Zr & 13.836 & 112.76 \\
Zr--O2--Zr & 13.836 & 112.76 \\
Zr--O2--H1 & 27.681 & 105.65 \\
O1--Zr--D  & 13.836 & 78.872 \\
O1--Zr--D $^{*}$  & 13.836 & 141.30 \\
O2--Zr--D  & 13.836 & 78.872 \\
O2--Zr--D $^{*}$  & 13.836 & 141.30 \\
D--Zr--D   & 150.0  & 77.304 \\
D--Zr--D $^{*}$  & 150.0  & 124.086 \\
O3--C1--O3 & 145.0  & 125.0 \\
O3--C1--C2 & 54.495 & 117.3 \\
C1--C2--C3 & 34.680 & 120.0 \\
C3--C2--C3 & 90.0   & 120.0 \\
C2--C3--C3 & 90.0   & 120.0 \\
C2--C3--H2 & 37.0   & 120.0 \\
C3--C3--H2 & 37.0   & 120.0 \\

\bottomrule
\end{tabularx}

$^{*}$Two distinct angular configurations are possible for these interactions due to the geometry of the coordination environment.

\end{table*}
\FloatBarrier
Table S1: nb-UiO-FF parameters part 1 out of 4.

\begin{table*}[!htbp]
\centering
\caption{Dihedral and improper parameters for UiO-66. }
\label{tab:dihedral_improper}
\begin{tabularx}{\textwidth}{X X X X}
\toprule
\midrule

\textbf{Dihedrals} & \textbf{$K$ [kcal/mol]} & \textbf{$d$} & \textbf{$n$} \\
\midrule
O3--C1--C2--C3 & 3 & $-1$ & 2 \\
C1--C2--C3--C3 & 3 & $-1$ & 2 \\
C1--C2--C3--H2 & 3 & $-1$ & 2 \\
C2--C3--C3--C2 & 3 & $-1$ & 2 \\
C2--C3--C3--H2 & 3 & $-1$ & 2 \\
C3--C3--C2--C3 & 3 & $-1$ & 2 \\
H2--C3--C2--C3 & 3 & $-1$ & 2 \\
H2--C3--C3--H2 & 3 & $-1$ & 2 \\

\midrule
\textbf{Impropers} & \textbf{$K$ [kcal/mol]} & \textbf{$d$} & \textbf{$n$} \\
\midrule
C2--C1--O3--O3 & 10.0  & $-1$ & 2 \\
C1--C2--C3--C3 & 10.0  & $-1$ & 2 \\
C2--C3--C3--H2 & 0.370 & $-1$ & 2 \\

\bottomrule
\end{tabularx}
\end{table*}

Table S2: nb-UiO-FF parameters part 2 out of 4.
\FloatBarrier

\begin{table*}[!htbp]
    \centering
    \caption{Non-bonded force field parameters for UiO-66.}
    \label{tab:charges_lj}
    
    \begin{tabularx}{\textwidth}{X X X X}
    \toprule
    \textbf{Atom type} & \textbf{Charge [e]} & \textbf{$\sigma$ [\AA]} & \textbf{$\varepsilon$ [kcal/mol]} \\
    \midrule

    Zr  &  0        &   2.783   &   0.069   \\
    O1  & -0.744    &   3.118   &   0.06    \\
    O2  & -0.877    &   3.118   &   0.06    \\
    H1  &  0.133    &   2.846   &   0.0152  \\
    D   &  0.492    &   0.0     &   0.0001  \\
    C1  &  0.630    &   3.473   &   0.0951  \\
    C2  & -0.082    &   3.473   &   0.0951  \\
    C3  & -0.065    &   3.473   &   0.0951  \\
    O3  & -0.586    &   3.118   &   0.06    \\
    H3  &  0.133    &   2.846   &   0.0152  \\

    \bottomrule
    \end{tabularx}
\end{table*}

\FloatBarrier

Table S3: nb-UiO-FF parameters part 3 out of 4.

\begin{table*}[!htbp]
    \centering
    \caption{Morse potential parameters for the interaction between Zr and O3, which models node/ligand reactivity.}
    \label{tab:morse}
    
    \begin{tabularx}{\textwidth}{X X X X}
    \toprule
    \textbf{Atom pair} & \textbf{$D_0$ [kcal/mol]} & \textbf{$\alpha$ [$\text{\AA}^{-1}$]} & \textbf{$r_0$ [\AA]} \\
    \midrule

    Zr--O3 &  1.3  &  4.1  &  2.25  \\

    \bottomrule
    \end{tabularx}
\end{table*}

Table S4: nb-UiO-FF parameters part 4 out of 4.

\begin{figure}[h!]
    \centering
    \noindent
    \begin{subfigure}[t]{0.48\columnwidth}
        \raggedright (a)
        \centering
        \includegraphics[trim=70px 70px 70px 70px, clip, width=\linewidth, keepaspectratio]{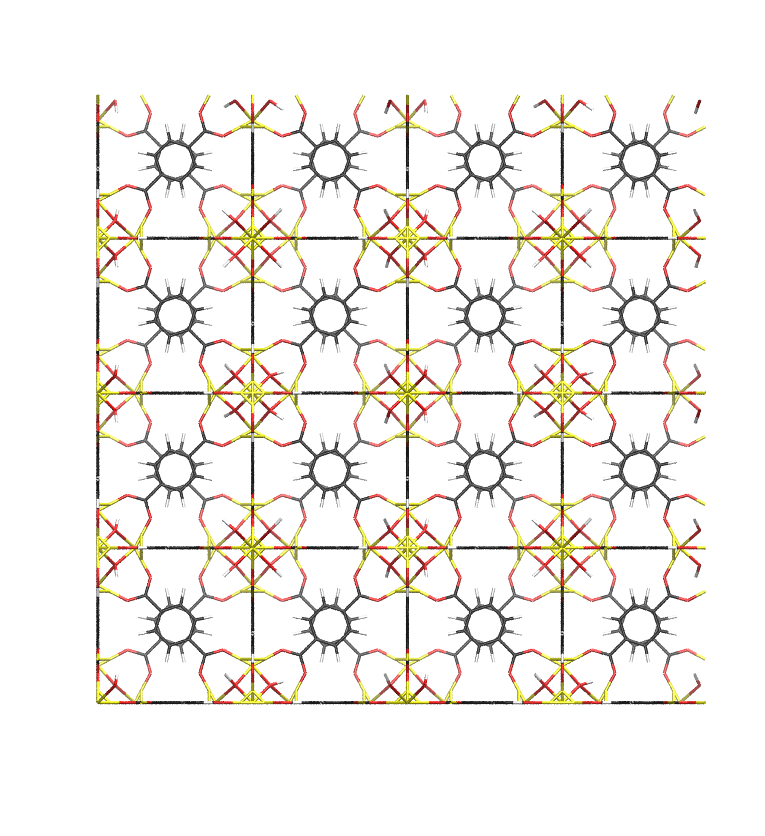}
        \label{fig:dummy_bulk}
    \end{subfigure}
    \hfill
    \begin{subfigure}[t]{0.48\columnwidth}
        \raggedright (b)
        \centering
        \includegraphics[trim=350px 10px 350px 10px, clip, width=\linewidth, keepaspectratio]{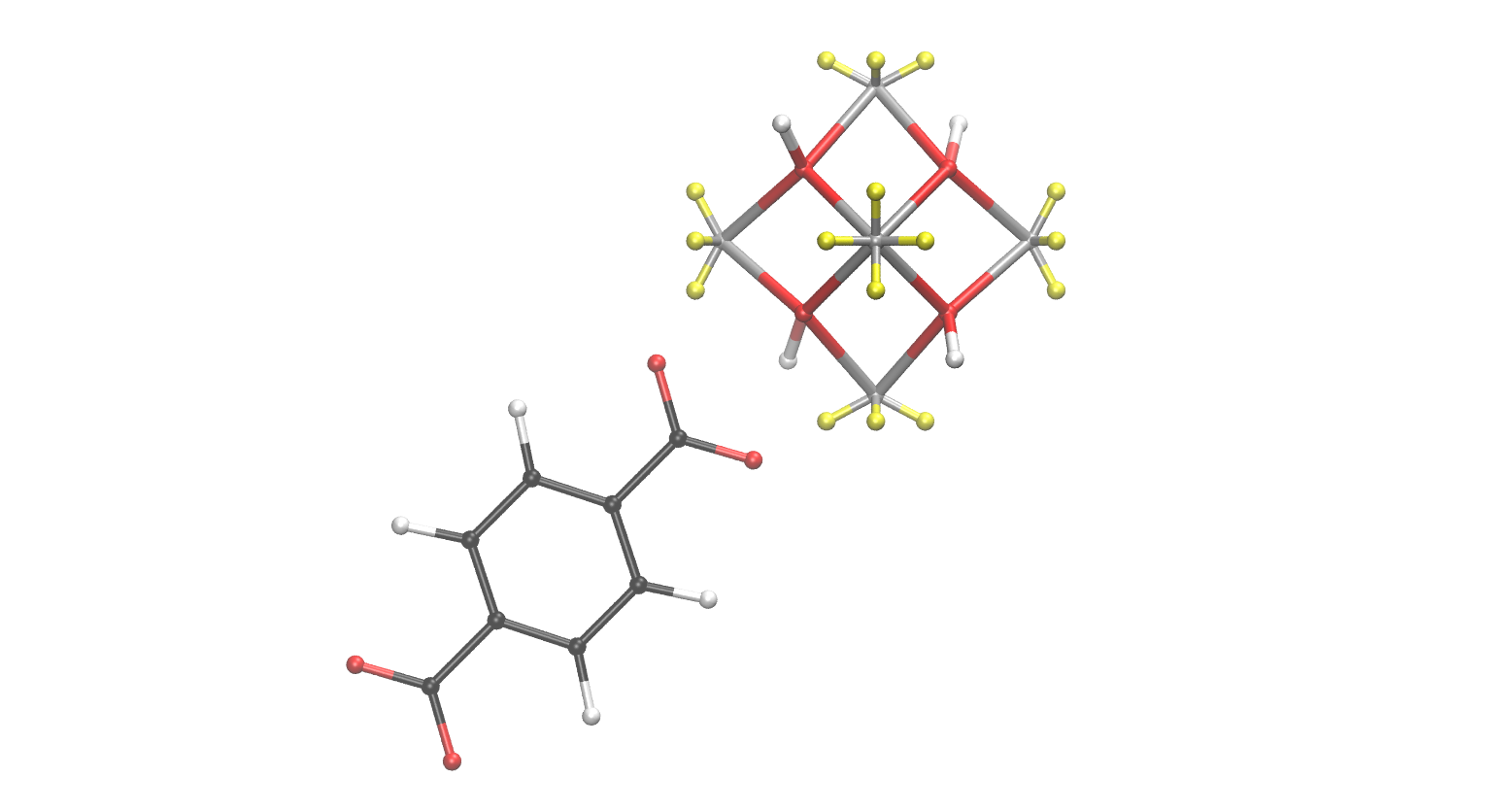}
        \label{fig:dummy_node}
    \end{subfigure}
    
    \caption{Visual representation of dummy atom placement within the UiO-66 framework. (a) UiO-66 2$\times$2$\times$2 supercell. Color scheme is the same as in Fig.~S1. (b) Dummy atoms shown around one node and a ligand oriented in the direction in which it will bind.}
    \label{fig:dummy_atoms_main}
\end{figure}

Figure S2: Illustration of the UiO-66 components and crystal structure including dummy atoms to help the reader understand the force field.

\begin{figure}[htbp]
    \centering
    \includegraphics[width=0.33\linewidth]{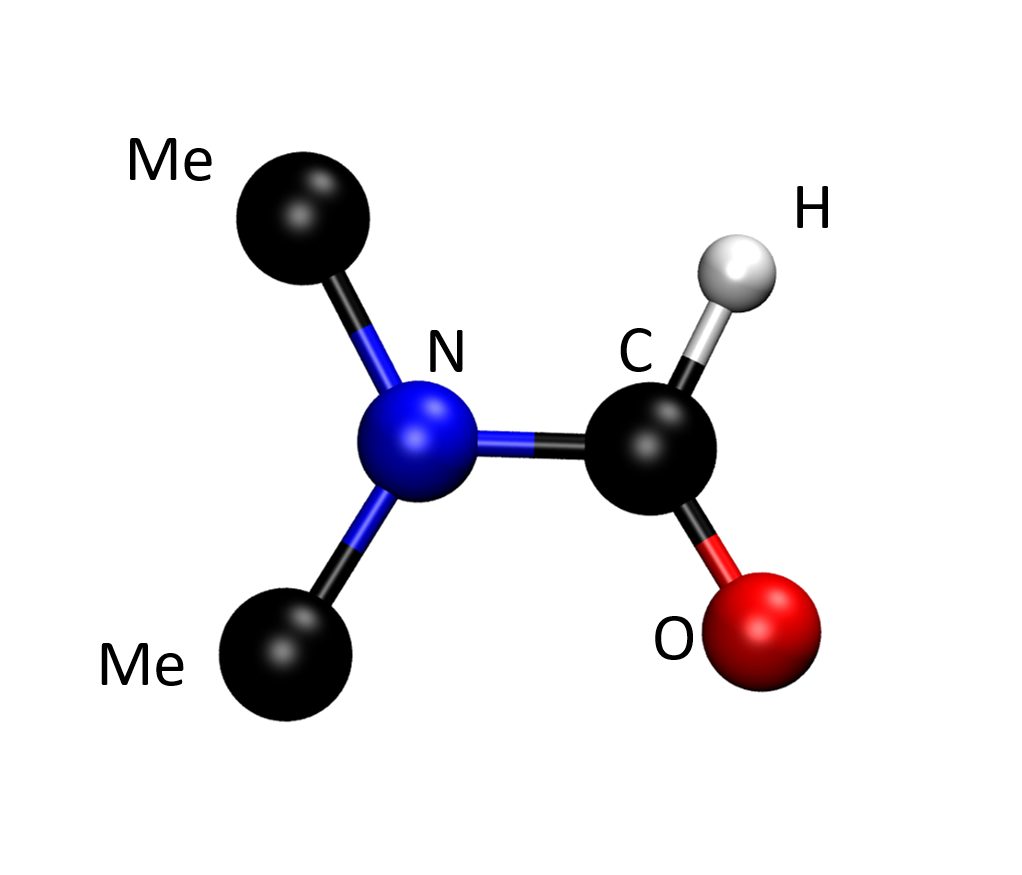}
    \caption{Definition of atom types used in the force field for DMF. Atoms are labeled according to their chemical environment. Me stands for methyle (united atom force field). Color code: O (red), H (white), C/Me (black), N (blue)}
    \label{fig:atom_types}
\end{figure}

\FloatBarrier

Figure S3: Illustration of atom types for DMF, these will be used in the force field tables.

\begin{table*}[htbp]
\centering
\caption{Bond stretching and angle bending parameters for DMF. Bond and angle interactions were described using harmonic potentials, $E_{\mathrm{bond}}=K_b(r-r_0)^2$ and $E_{\mathrm{angle}}=K_\theta(\theta-\theta_0)^2$, respectively.}
\label{tab:dmf_bond_angles} 
\begin{tabularx}{\textwidth}{X X X}
\toprule
\midrule

\textbf{Bonds} & \textbf{$K_b$ [kcal/(mol$\cdot$\AA$^2$)]} & \textbf{$r_0$ [\AA]} \\
\midrule
C--H  & 316.852 & 1.123 \\
C--O  & 650.077 & 1.23 \\
C--N & 430.079 & 1.33 \\
N--Me & 240.06 & 1.44   \\

\midrule

\textbf{Angles} & \textbf{$K_{\theta}$ [kcal/(mol$\cdot$rad$^2$)]} & \textbf{$\theta_0$ [deg]} \\
\midrule
Me--N--Me & 50.041 & 121.0 \\
Me--N--C & 50.041 & 119.5 \\
N--C--H & 43.976 & 114.5 \\
O--C--N & 74.947 & 123.0 \\
H--C--O & 43.976 & 122.5 \\

\bottomrule
\end{tabularx}
\end{table*}

Table S5: DMF force field part 1 out of 3.

\begin{table*}[!htbp]
\centering
\caption{Dihedral parameters for DMF. The dihedral interactions were described using a periodic potential of the form $E_{\mathrm{dihedral}}=K[1+d\cos(n\phi)]$.}
\label{tab:dmf_dihedral_improper}
\begin{tabularx}{\textwidth}{X X X X}
\toprule
\midrule

\textbf{Dihedrals} & \textbf{$K$ [kcal/mol]} & \textbf{$d$} & \textbf{$n$} \\
\midrule
Me--N--C--H & 2.7673 & $-1$ & 2 \\
Me--N--C--O & 2.7673 & $-1$ & 2 \\

\bottomrule
\end{tabularx}
\end{table*}

Table S6: DMF force field part 2 out of 3.

\begin{table*}[!htbp]
    \centering
    \caption{Non-bonded force field parameters for DMF.}
    \label{tab:dmf_charges_lj}
    
    \begin{tabularx}{\textwidth}{X X X X}
    \toprule
    \textbf{Atom type} & \textbf{Charge [e]} & \textbf{$\sigma$ [\AA]} & \textbf{$\varepsilon$ [kcal/mol]} \\
    \midrule

    N  & -0.57    &   3.2   &   0.1555     \\
    C  &  0.45    &   3.7   &   0.0972     \\
    O  & -0.5     &   2.96  &   0.2183    \\
    H  &  0.06    &   2.2   &   0.0155     \\
    Me &  0.28    &   3.79  &   0.0155    \\

    \bottomrule
    \end{tabularx}
\end{table*}

Table S7: DMF force field part 3 out of 3.

\begin{figure}[htbp]
    \centering
    \includegraphics[width=0.33\linewidth]{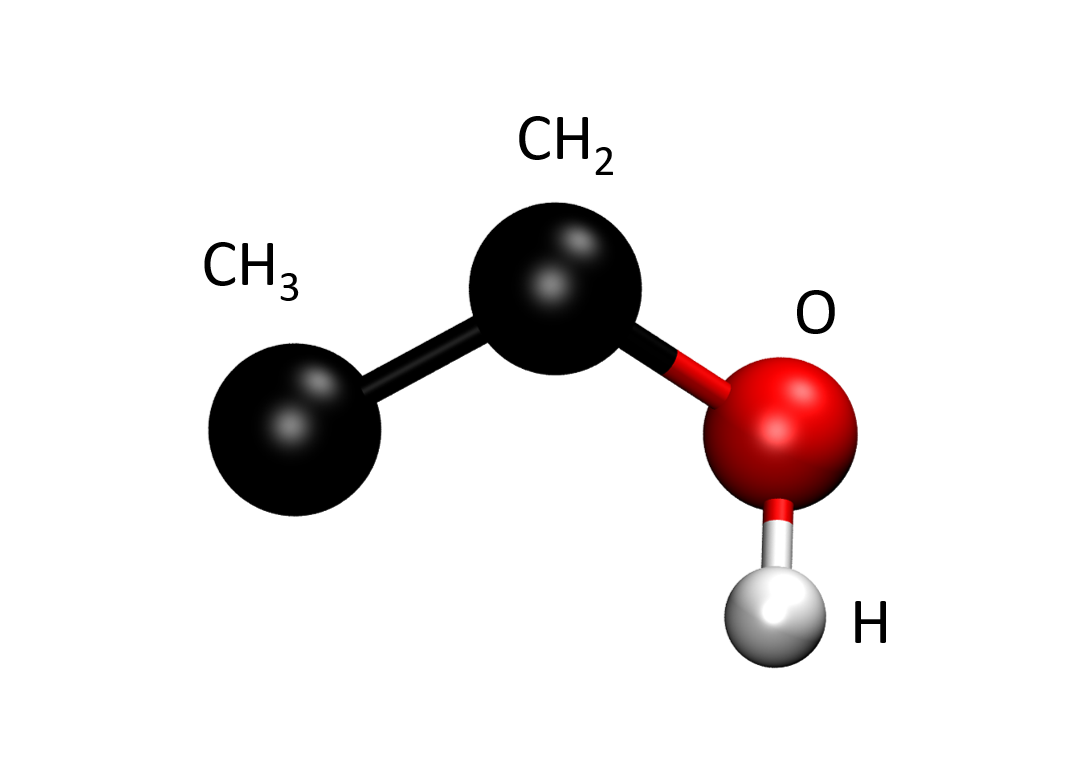}
    \caption{Definition of atom types used in the force field for ethanol. Atoms are labeled according to their chemical environment. Color code: O (red), H (white), $CH_3$/$CH_2$ (black)}
    \label{fig:atom_types}
\end{figure}

\FloatBarrier

Figure S4: Illustration of atom types for ethanol, these will be used in the force field tables.

\begin{table*}[htbp]
\centering
\caption{Bond stretching and angle bending parameters for ethanol. Bond and angle interactions were described using harmonic potentials, $E_{\mathrm{bond}}=K_b(r-r_0)^2$ and $E_{\mathrm{angle}}=K_\theta(\theta-\theta_0)^2$, respectively.}
\label{tab:eth_bond_angles} 
\begin{tabularx}{\textwidth}{X X X}
\toprule
\midrule

\textbf{Bonds} & \textbf{$K_b$ [kcal/(mol$\cdot$\AA$^2$)]} & \textbf{$r_0$ [\AA]} \\
\midrule
$CH_3$--$CH2$  & 116.26  & 1.54  \\
$CH_2$--O      & 146.7   & 1.43  \\
O--H           & 281.755 & 0.945 \\

\midrule

\textbf{Angles} & \textbf{$K_{\theta}$ [kcal/(mol$\cdot$rad$^2$)]} & \textbf{$\theta_0$ [deg]} \\
\midrule
$CH_2$--O--H      & 110.090 & 108.5 \\
$CH_3$--$CH_2$--O & 100.154 & 109.5 \\

\bottomrule
\end{tabularx}
\end{table*}

Table S8: Ethanol force field part 1 out of 3.

\begin{table*}[!htbp]
\centering
\caption{Dihedral parameters for ethanol. The torsional interactions were described using the OPLS Fourier series potential, $E_{\mathrm{dihedral}}=\frac{1}{2}K_1[1+\cos(\phi)]+\frac{1}{2}K_2[1-\cos(2\phi)]+\frac{1}{2}K_3[1+\cos(3\phi)]+\frac{1}{2}K_4[1-\cos(4\phi)]$.}
\label{tab:eth_dihedral_improper}
\begin{tabularx}{\textwidth}{X X X X X}
\toprule
\midrule

\textbf{Dihedrals} & \textbf{$K_1$ [kcal/mol]} & \textbf{$K_2$ [kcal/mol]} & \textbf{$K_3$ [kcal/mol]} & \textbf{$K_4$ [kcal/mol]}\\
\midrule
$CH_3$--$CH_2$--O--H & 0.8339 & -0.1159 & 0.7469 & 0 \\

\bottomrule
\end{tabularx}
\end{table*}

Table S9: Ethanol force field part 2 out of 3.

\begin{table*}[!htbp]
    \centering
    \caption{Non-bonded force field parameters for ethanol.}
    \label{tab:eth_charges_lj}
    
    \begin{tabularx}{\textwidth}{X X X X}
    \toprule
    \textbf{Atom type} & \textbf{Charge [e]} & \textbf{$\sigma$ [\AA]} & \textbf{$\varepsilon$ [kcal/mol]} \\
    \midrule

    $CH_3$  &  0.265    &   3.95    &   0.0914       \\
    $CH_2$  &  0.0      &   3.75    &   0.195        \\
    O       & -0.7      &   3.02    &   0.185        \\
    H       &  0.435    &   0.00    &   0.00001      \\

    \bottomrule
    \end{tabularx}

    \small
    \textit{Note:} Cross interactions were generated using Lorentz--Berthelot mixing rules. Cross interactions for H--O1 / H--O2 (oxygen atoms of the UiO-66 node) were modified by changing the $\sigma$ value from 1.559 to 2.027~\AA{} (30\% increase).
\end{table*}

\FloatBarrier

Table S10: Ethanol force field part 3 out of 3.

\FloatBarrier
\newpage

\section{Additional Results}

\begin{figure*}[h!]
     \centering
     \begin{subfigure}{0.4\textwidth}
         \centering
         \includegraphics[width=\textwidth]{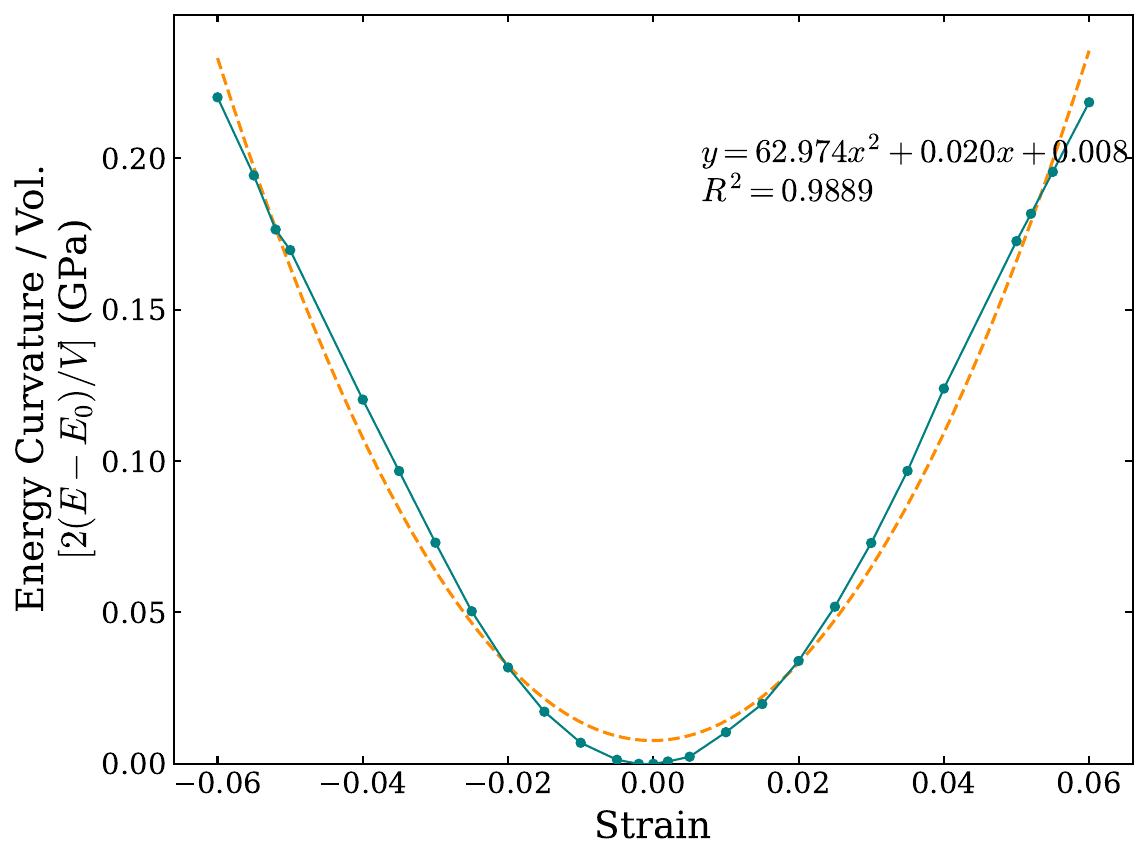}
         \label{fig:c11}
     \end{subfigure}
     
     \vspace{0.3cm} 

     \begin{subfigure}{0.4\textwidth}
         \centering
         \includegraphics[width=\textwidth]{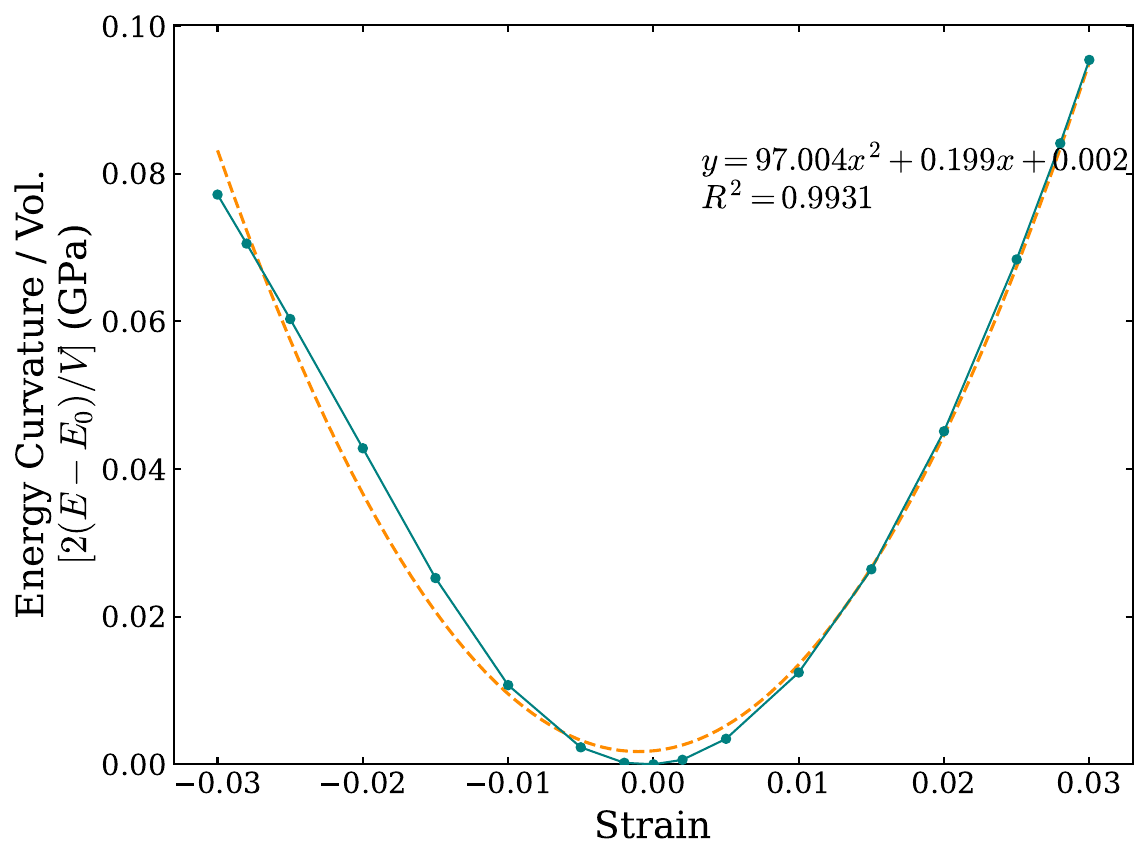}
         \label{fig:c12}
     \end{subfigure}

     \vspace{0.3cm}

     \begin{subfigure}{0.4\textwidth}
         \centering
         \includegraphics[width=\textwidth]{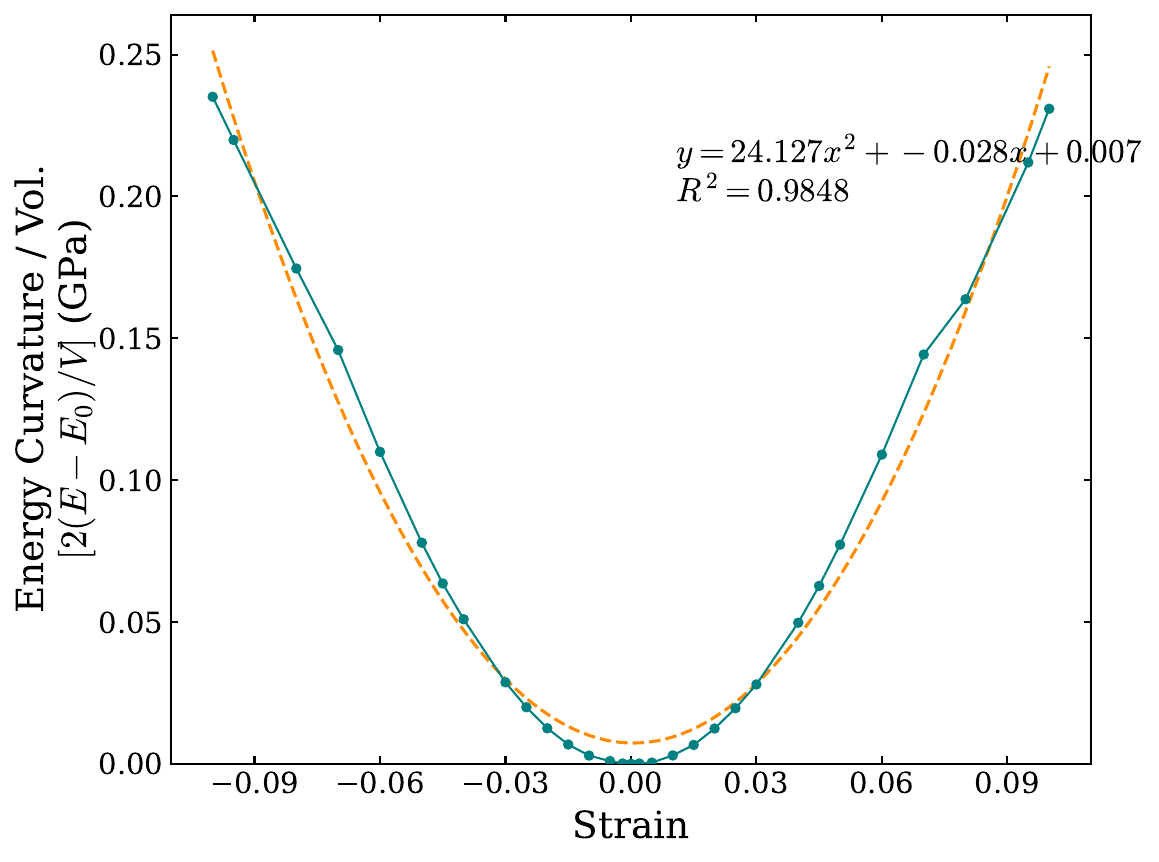}
         \label{fig:c44}
     \end{subfigure}
     
     \caption{Elastic constants fits for the nb-UiO-FF framework: (a) C11 (b) C22 and (c) C44.}
     \label{fig:elastic_consts}
\end{figure*}

Figure S5: Plots from which elastic constants are obtained. The purpose is to show the quality of the estimation.

\begin{table*}[!htbp]
    \centering
    \caption{Mechanical properties of UiO-66 calculated from its elastic constants using nb-UiO-FF.}
    \label{tab:mecha}
    
    \begin{tabularx}{\textwidth}{X X X X}
    \toprule
    \textbf{Mechanical Property} & \textbf{Analytical Relation} & \textbf{Value (GPa)} \\
    \midrule
    Young's Modulus ($E$) & $\displaystyle \frac{(C_{11}-C_{12})(C_{11}+2C_{12})}{C_{11}+C_{12}}$ & 40 \\ 
    Poisson's Ratio ($\mu$) & $\displaystyle \frac{C_{12}}{C_{11}+C_{12}}$ & 0.35 \\ 
    Bulk Modulus ($B$) & $\displaystyle \frac{1}{3} (C_{11}+2C_{12})$ & 44 \\ 
    \bottomrule
    \end{tabularx}
\end{table*}

\FloatBarrier

Table S11: Values of mechanical properties computed using the elastic constants obtained from the force field.  

\begin{figure}[h!]
    \centering

    \begin{subfigure}[t]{0.55\textwidth}
        \centering
        \includegraphics[width=\linewidth]{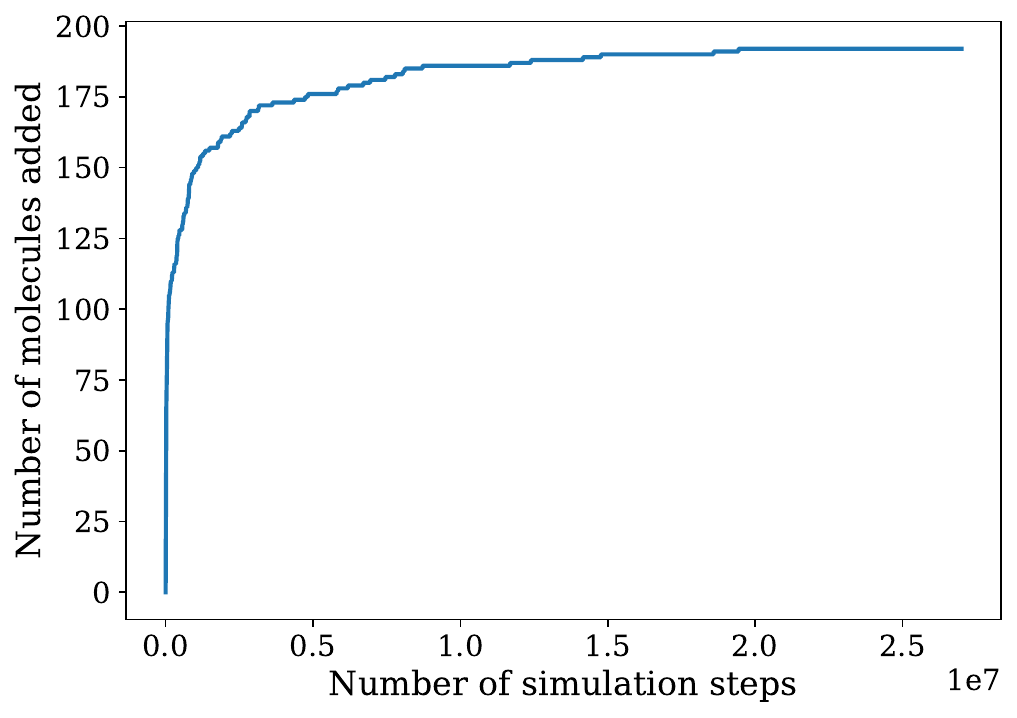}
        \caption{A total of 192 DMF molecules were added.}
        \label{fig:gcmc_dmf}
    \end{subfigure}
    \hfill

    \begin{subfigure}[h!]{0.55\textwidth}
        \centering
        \includegraphics[width=\linewidth]{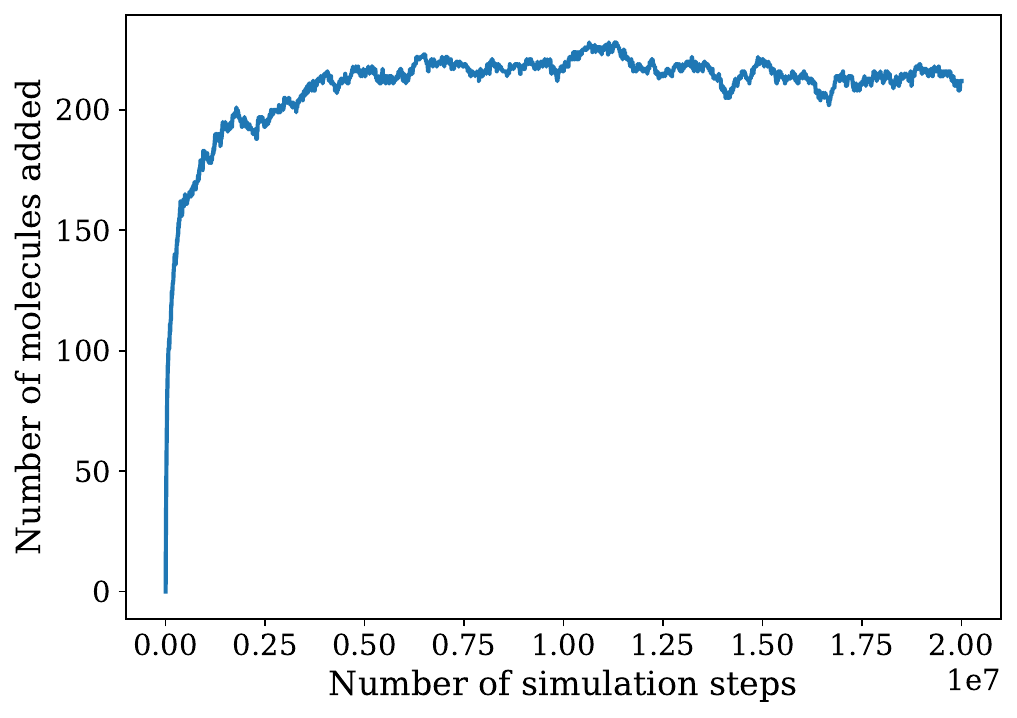}
        \caption{A total of 212 Ethanol molecules were added.}
        \label{fig:gcmc_ethanol}
    \end{subfigure}

    \caption{Evolution of the number of adsorbed molecules over time in a 2$\times$2$\times$2 UiO-66 supercell for two different solvents from Grand Canonical Monte Carlo simulations. The plateau indicates convergence to equilibrium loading. (a) DMF, (b) Ethanol.}
    \label{fig:gcmc_uptake}
\end{figure}

\FloatBarrier

Figure S6: Time evolution of solvent loadings in UiO-66 computed with our force field. The purpose is to show that we reached the saturation value, and to obtain that value. The number of molecules increases as the simulation progresses and plateaues at the saturation point.

\begin{figure}[h!]
    \centering
    \includegraphics[width=0.55\linewidth]{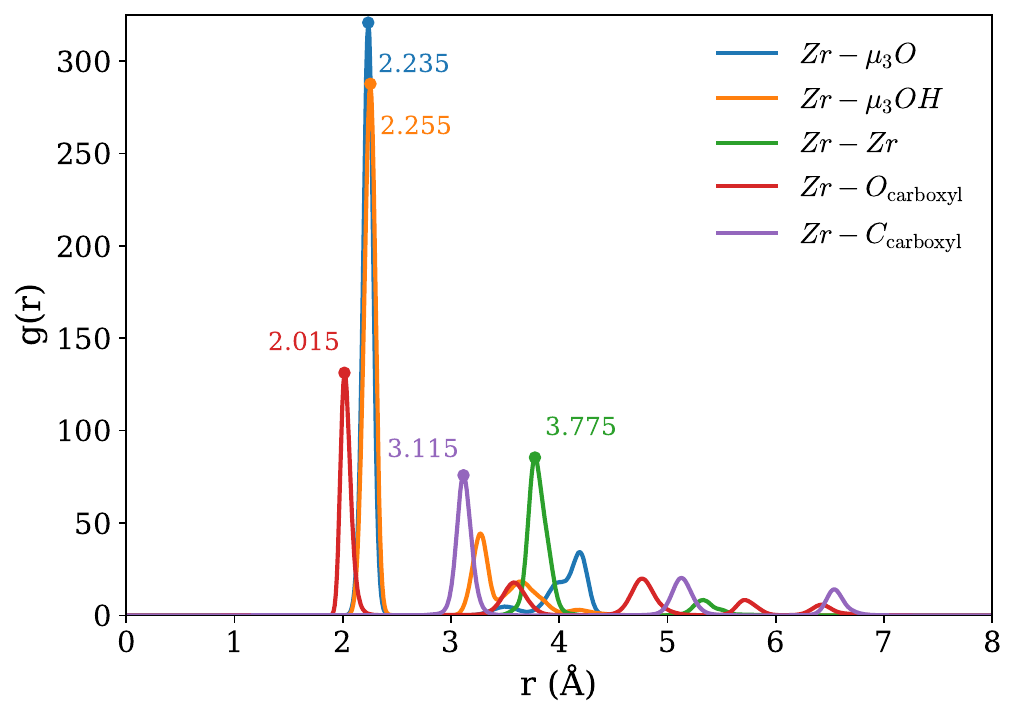}
    \caption{Radial distribution functions for selected types in UiO-67.}
    \label{fig:RDFs_UiO-67}
\end{figure}

\FloatBarrier

Figure S7: Radial distribution functions plots for UiO-67, these provide structural details including bond lengths for this MOF.

\begin{figure}[h!]
    \centering
    \includegraphics[width=0.55\linewidth]{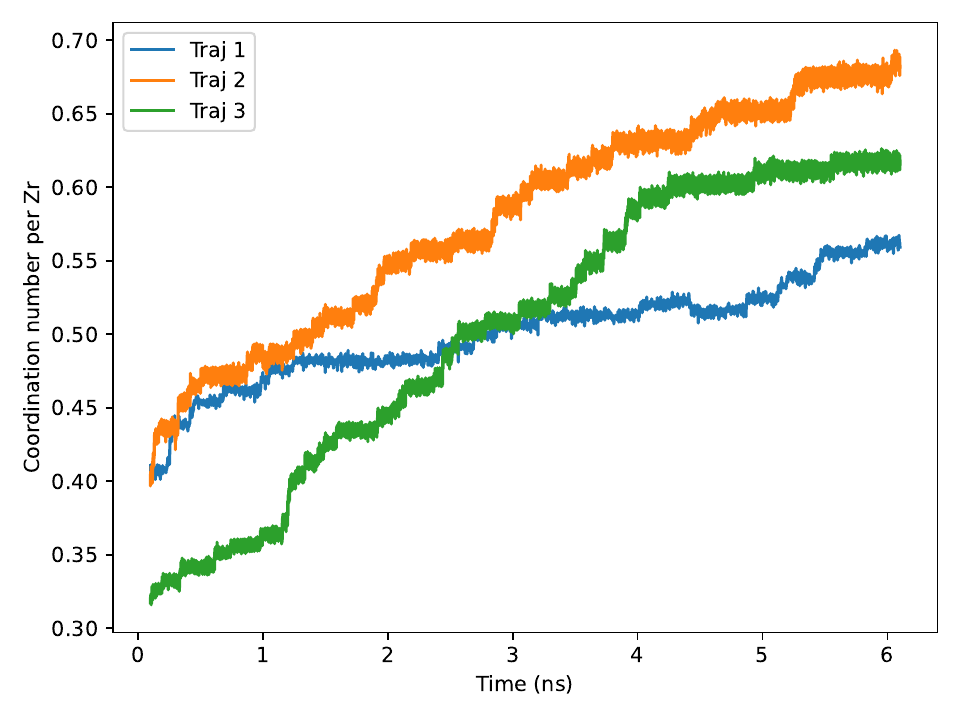}
    \caption{Time evolution of the coordination number per Zr atom for the three independent self-assembly simulation trajectories.}
    \label{fig:coordination}
\end{figure}

Figure S8: Time evolution of the Zr--O coordination number. The purpose is to show that it increases over time, indicating that self-assembly has begun and progresses with time.

Coordination number was evaluated using a continuous switching function of the rational form implemented in PLUMED\cite{TRIBELLO2014604} given by \( s(r)=1-(r-d_0)^{n}/[1-(r-d_0)^{m}] \). A cutoff distance corresponding to the $\ce{Zr}-\ce{O_{ligand}}$ bond length {$R_0$} = 2.3 \AA \ was employed. The parameters of the switching function were set to n=6, m=0, and $d_0$=0.

\end{document}